\documentstyle[preprint,aps]{revtex}
\begin{document}
\tightenlines
\pagestyle{empty} 
\centerline{\hfill             CERN--TH.6839/93}
\centerline{\hfill                NUHEP-TH-93-9}
\centerline{\hfill               hep-ph/9307232}
\centerline{\Large 
CP Violating Observables in $e^-e^+\rightarrow W^-W^+$}
\vskip 1cm
\centerline{
Darwin Chang$^{(1)}$,
Wai--Yee Keung$^{(2,3)}$ and
Ivan Phillips$^{(1)}$
}
\centerline{
$^{(1)}$Department of Physics and Astronomy,}
\centerline{Northwestern University, Evanston, IL 60208 }
\vskip .3cm
\centerline{$^{(2)}$Theory Group, CERN
CH-1211, Geneva 23, Switzerland}
\vskip .3cm
\centerline{$^{(3)}$Physics Department, University of Illinois 
at Chicago, IL 60607--7059}

\vskip 1.5cm
\centerline{ABSTRACT}
\vskip .5cm
We consider various integrated lepton charge-energy asymmetries and
azimuthal asymmetries as tests of CP violation in the process $e^-e^+
\rightarrow W^-W^+$. These asymmetries are sensitive to different linear
combinations of the CP violating form factors in the three gauge boson
$W^-W^+$ production vertex, and can distinguish dispersive and
absorptive parts of the form factors.  It makes use of purely hadronic
and purely leptonic modes of $W$'s decays as well as the mixed modes. 
The techniques of using the kinematics of jets or missing momentum to
construct CP--odd observables are also employed. These CP violating
observables are illustrated in the generalized Left-Right Model and the
Charged Higgs Model. 
\vskip 1cm

\noindent
PACS numbers: 11.30.Er, 14.80.Er
\vskip 1cm
\centerline{Published in Phys. Rev. D{\bf 48}, 4045-4061 (1993).}
\eject
\narrowtext
\pagestyle{plain}
\section{INTRODUCTION}
\label{sec:intro}
 
Since the discovery\cite{cronin} of CP violation (CPV) in 1964, a satisfactory 
understanding of its origin has eluded us.  One reason for this
is that there is very little experimental information about CPV.
The measurement of the $\epsilon$ parameter in the $K^0-\overline{K^0}$ 
system remains the only evidence of CPV.  
The searches for CP violating effects in other physical
systems (such as the $\epsilon'$ parameter or 
the electric dipole moments of neutron or electron) 
have produced only negative constraints.
The detection of new CP violating effects will certainly 
greatly enhance our understanding of this phenomenon.  
One potential stage for such new effects are high energy 
collisions in existing or future colliders.  This is an exciting 
possibility because it is well-known that within the Standard 
Kobayashi--Maskawa (KM) Models\cite{KM} the possibility of detecting 
CPV in high energy collisions is very small.  Therefore, any evidence 
we can record in this arena will be a window into the physics beyond the 
Standard Model.
 
In this paper we shall explore the possibility of detecting CPV in the 
production of $W^+W^-$ pairs in the leptonic colliders.  Among the various 
interactions that can contribute to this process, the vertices associated 
with the couplings of $W$ gauge bosons to photons or $Z$ bosons have not been  
probed strongly by the present experimental data.  
Therefore we shall consider these vertices as the potential sources of the 
new CP violating phenomena.  
One can start by writing down the most general three vector
boson coupling\cite{hagiwara,ref:Gaemers} consistent with gauge and 
Lorentz invariance.  It was found that\cite{hagiwara,ref:Gaemers} 
the effective vertices for $W^+W^-V$, where $V$ is either a photon or $Z$ boson,
may be parametrized in terms of seven form factors for each V, three
linear combinations for each case are CP violating.  
 
These CP violating form factors 
are zero even at one loop level in the KM Model.
However, when we introduce new physics such as Left-Right Symmetry,
Supersymmetry, or additional Higgs multiplets, CP violation can appear at one
loop and lead to significant CP non-conservation expressed in the three gauge
boson vertex.  However, one should always keep in mind that CPV models using 
renormalizable gauge theories are just simple examples of what can happen 
beyond the Standard Model.  Since CP is such a fundamental symmetry, the fact 
that one can probe these form factors is already very interesting even if 
one can not produce a renormalizable gauge theory that gives large 
form factors.
The LEP--II collider at CERN and its potential upgrades are ideal for 
studying the properties of the three gauge boson vertices including CPV.
 
The pair production process $e^-e^+ \rightarrow x \bar x$, where $x$ is some
particle, is characterized only by the angle between the $x$ momentum and the
$e^-$ momentum  in the CM frame, and the helicities of the particles.
If the particle helicities are averaged over, one can easily show that 
the process is
C, P and CP self-symmetric. For this reason the helicities need to be
determined or statistically analyzed to observe violations of
these discrete symmetries in the pair production process.
Here we shall assume that the electron and positron beam are unpolarized.  
We shall employ the dependence of final state momenta on the helicity of 
the decaying $W$'s to probe CP property. 
 
There are basically two different types of helicity analyzers. The
leptonic decay of the $W$ is an example of the first type. 
A $W^-$ boson at rest with its spin along the $z$-axis preferentially
emits a lepton in the $-z$ direction.  When we boost the $W^-$ in the
$+z$-direction, the leptons $\ell^-$ from the $W^-$ with positive
helicity are on average softer than those with zero or negative
helicities. Hence, if there is a CP violating asymmetry in the helicity
states of the $W^-W^+$ pair, it may show up as a difference in the
number of positive and negative leptons in each lepton energy bin. This
is investigated in Section III. A second way to get information about
the  helicity is to zoom into the azimuthal angular dependence of the
outgoing fermions. This can be done easily for the outgoing charged
leptons.  When the the outgoing fermions are quarks, one has to
look at the relative azimuthal angular dependence of the decay planes of
the two $W$ bosons without sensitivity to the charge of the
associated $W$ bosons; this also allows detection of CP violation at
the production vertex as we show in Section IV.  However, for the
purpose of detecting CP violating signals whatever the source is, it
may be more natural to start from kinematic variables that are
easier to detect.  Some examples of this are emphasized in Section
V and VI.
 
CP violating effects in $e^-e^+ \rightarrow W^-W^+$ have been analyzed 
before \cite{hagiwara,gounaris,othereewwcp}.  
In this article we make a comprehensive analysis of 
some integrated observables which are also 
sensitive to all of the parameters of the
gauge boson vertices.  
Some of these are just the integrated or partially-integrated versions of 
the asymmetries considered in Ref.\cite{gounaris}.  Some of them were never
considered before.  
These new observables are somewhat more intuitive
than the weighting approach of Ref.\cite{gounaris}.  They also allow one to 
make use of all the decay modes of $W$ pairs.  The process by
which these observables are constructed is quite general and may be
applied to other
systems\cite{r:cphiggs,r:hcpodd,r:eett,r:othereett,r:CPjet}.
In section \ref{sec:couplings}, we describe
the effective $ZWW$ and $\gamma WW$ vertices.
In section III, we show the energy asymmetry between the lepton and the
anti--lepton from the decay of $W^\pm$.
In section IV, we setup a framework to study the CP--odd, 
CP$\hat{\hbox{T}}$-even angular asymmetry.  
We further pursue other CP violating observables in
section V using the  purely hadronic mode, 
and section VI (purely leptonic mode).
Models which give
significant CP violation in the vertices are discussed in section
\ref{sec:models}.
\section{THE THREE GAUGE BOSON VERTEX AND HELICITY AMPLITUDES}
\label{sec:couplings}
 
Hagiwara {\it et al.} have made a detailed analysis\cite{hagiwara} of
the $WW$ production processes.  In this section we shall give a summary
of the general operator structure involved. The three gauge boson
effective vertex for a vector boson $V$ coupling to two $W$ bosons is
\begin{eqnarray}
\Gamma^{\alpha\beta\mu}_{V^*(P)\to W^-(q)W^+(\bar q)}
            &=& f_1^V(q-\bar q)^{\mu} g^{\alpha\beta}
              - f_2^V(q-\bar q)^{\mu} P^{\alpha}P^{\beta}  \nonumber \\
 &+&f_3^V(P^{\alpha} g^{\mu \beta} - P^{\beta} g^{\mu \alpha})
   +if_4^V(P^{\alpha} g^{\mu \beta} + P^{\beta} g^{\mu \alpha})
\nonumber\\
 &+&if_5^V \varepsilon^{\mu \alpha \beta \rho} (q-\bar q)_{\rho}
    -f_6^V \varepsilon^{\mu \alpha \beta \rho} P_{\rho} \nonumber \\
 &-& ({f_7^V}/{m_W^2}) (q-\bar q)^{\mu} \varepsilon^{\alpha \beta \rho \sigma}
   P_{\rho} (q-\bar q)_{\sigma}.
\end{eqnarray}
These are the most general operator structure for on-shell $W$ bosons.
In the above, $P$ is the momentum of the vector particle $V$ into the vertex,
and $q$ and $\bar q$ are the outgoing momenta of the $W^-$ and $W^+$
respectively.  
$\alpha$ and $\beta$ are the polarization indices of the $W^-$ and $W^+$
respectively.  
All seven form factors, $f_1$--$f_7$, can have
absorptive and dispersive parts depending on the model and the kinematics.
The Standard Model predicts, at tree level,
\begin{equation}
f_1=1  \;,\qquad f_3=2\;, \qquad f_{2,4,5,6,7}=0 \;.
\end{equation}
Form factors $f_1$,$f_2$,$f_3$
and $f_5$ are CP even couplings 
while $f_4$, $f_6$, and $f_7$ are
CP odd.  Our first task is to decode both the real and the
imaginary parts of these form factors $f_{4,6,7}$ from
the CP violating observables constructed out of the asymmetries in the
scattering kinematics.
 
The second task is to identify potential CP violation renormalizable gauge 
models in which these form factors are induced significantly.  
In Standard KM Model, contributions to these form factors are very small 
because they are not generated until two or three 
loop level and they come with the typical light quark mass and 
mixing angle suppression factors.
In non-standard models, $f_4$ and $f_6$ are easily generated at one loop.
Gauge models are discussed in Section \ref{sec:models}.

When investigating the asymmetries induced by these form factors, one 
should be aware of the constraint imposed by the CPT theorem and 
unitarity condition.  
It is well-known that it is very difficult to check directly the effect 
the time reversal symmetry, T, or the CPT symmetry in process such as ours.  
This is because both symmetries require one to interchange the initial and 
the final states.   However, in the Born 
approximation, the unitarity of the $S$-matrix implies that the 
transition matrix, $M$, is hermitian, and this allows one to flip the 
initial state into the final state as long as the amplitude is also 
complex conjugated.  
We shall define a pseudo time reversal symmetry, 
which we shall call $\hat{\hbox{T}}$, that 
transforms only the kinematic observables of both the initial and final 
states according to time reversal but does not transform the initial and 
final state into each other as required by time reversal.  
Therefore as long as the transition matrix is hermitian, which is 
always the case in Born approximation, the CPT constraint can be reduced to 
\begin{equation}
\langle f| M |i\rangle = 
\langle \hbox{CP}\hat{\hbox{T}}(f)| M | \hbox{CP}\hat{\hbox{T}}(i)
\rangle^*,
\label{eq:cpt}
\end{equation}
where $|\hbox{CP}\hat{\hbox{T}}(j)\rangle$ represents the state 
$|j\rangle$ with it kinematic variables and quantum number numbers 
transformed by CP$\hat{\hbox{T}}$.
We shall call the transformation from $\langle f| M |i\rangle$ into 
$\langle\hbox{CP}\hat{\hbox{T}}(f)| M | \hbox{CP}\hat{\hbox{T}}(i)
\rangle^*$ the CP$\hat{\hbox{T}}$\ transformation.
Therefore the CPT theorem allows us to detect whether the transition matrix is 
hermitian or not by looking at the same process with its kinematic variables 
transformed by CP$\hat{\hbox{T}}$.
The nonhermiticity of the transition matrix occurs when the contributions 
beyond Born approximation are included in which some intermediate states can 
be on-shell.  
Such contributions are traditionally called final state interactions(FSI),  
(even though initial state rescattering can also give rise to a similar 
effect).  
In our case, since the initial state is CP self conjugate, we can label 
our observables according to their CP and CP$\hat{\hbox{T}}$\ properties.  
Clearly if an observable is CP$\hat{\hbox{T}}$-odd, 
FSI will be required for its observation.  
Similarly, if it is CP$\hat{\hbox{T}}$-even, 
there is no need of FSI for its observation.
%
Note however that in renormalizable gauge theory 
the fact that one needs FSI does not always mean that the effect will contain
an additional suppression factor.
Since CP violating effects are usually
loop effects, it is possible to incorporate the CP violation and the
rescattering (FSI) effects into the same one loop diagram.
The loop diagram then
interferes with the tree level process to produce an observable result.
In the form factor approach which we are adopting,
nonhermiticity of the transition matrix, which is the hallmark of 
the final state rescattering effect, can be represented as the imaginary 
parts of the form factors.   Therefore, to measure both the real and 
the imaginary parts of the form factors directly one needs both 
CP--odd, CP$\hat{\hbox{T}}$-even and CP--odd, CP$\hat{\hbox{T}}$-odd 
observables.
 
Helicity amplitudes for the $W$ pair production are given in section 3.1 of
Ref.\cite{hagiwara}.  They can be written as
\begin{equation}
   {\cal M}_{\sigma,\bar \sigma;\lambda,\bar \lambda} (\Theta)=
\sqrt{2} \: e^2 \: 
\tilde{\cal M}_{\sigma,\bar \sigma;\lambda,\bar \lambda}
                          (\Theta)
\: d^{max(|\Delta\sigma|,|\Delta\lambda|)}_{\Delta \sigma,\Delta \lambda}
                          (\Theta),
\label{eq:prod}
\end{equation}
where $\sigma$,$\bar \sigma$($=\pm 1$) are the electron and positron
helicities, $\lambda$, $\bar \lambda$ are the $W^-$ and $W^+$ helicities
respectively.  The angle $\Theta$ is the angle between the electron and
the $W^-$. $\Delta \sigma = \frac{1}{2}(\sigma - \bar \sigma)$, and
$\Delta \lambda= \lambda -\bar \lambda$. The coordinate system is shown
in Fig.~1. Note that our definition differs from that of Ref
\cite{hagiwara} by a factor of $\varepsilon=\Delta\sigma(-1)^{\bar
\lambda}$; this convention leads to simpler CP properties\cite{gounaris}
 for the density matrices of the $W$'s in section
\ref{sec:uncorrelated}. 
To be complete, we list the relevant $d$ functions,
\begin{eqnarray}
d^2_{1,\pm2}(\Theta)&=&-d^2_{-1,\mp2}(\Theta)=
\pm\case1/2(1\pm\cos\Theta) \sin\Theta          ,        \nonumber\\
d^1_{1,\pm1}(\Theta)&=& d^1_{-1,\mp1}(\Theta)=
   \case1/2(1\pm\cos\Theta)                     ,                 \\
d^1_{1,0}(\Theta)&=&-d^1_{-1,0}(\Theta)
=-d^1_{0,1}(\Theta)=d^1_{0,-1}(\Theta)
=-{1 \over \sqrt{2}}\sin\Theta   .
\nonumber
\end{eqnarray}
Due to the angular momentum conservation, 
the amplitude is non--vanishing only when
$\bar \sigma =-\sigma$ in the high energy limit $\sqrt{s}\gg m_e$.  
Therefore only $\Delta \sigma = \pm 1$ case concerns us.
\begin{eqnarray}
\tilde{\cal M}_{\sigma,-\sigma;\lambda,\bar\lambda}&=&
{\beta \over \sin^2\theta_W}(- \case1/2 \delta_{\sigma,-1}+\sin^2\theta_W)
A^Z_{\lambda,\bar \lambda}{s \over s-m_Z^2}
                       -\beta A^\gamma_{\lambda,\bar \lambda}
                                                    \nonumber\\
    &+& \delta_{\sigma,-1}{1 \over 2\beta\sin^2\theta_W}
    [B_{\lambda,\bar \lambda} -
    {C_{\lambda,\bar \lambda} \over 1+\beta^2-2\beta \cos\Theta}]
\;,
\label{eq:amplitude}
\end{eqnarray}
with $\beta^2=1-\gamma^{-2}$, 
$\gamma=\case1/2\sqrt{s}/m_W$ and $\sin^2\theta_W=0.23$.
Coefficients $A^\gamma$ and $A^Z$ are related to amplitudes due to
$\gamma^*$, $Z^*$ in the $s$--channel.
Coefficients $B$ and $C$ are related to
the $t$--channel neutrino exchange diagram.
Note that, for $\Delta \lambda=\pm 2$,
$A_{\lambda,\bar\lambda}=0$ and
$B_{\lambda,\bar\lambda}=0$, only coefficients $C$ from the $\nu$
contribution survive.
We shall tabulate $A_{\lambda,\bar\lambda}$, $B_{\lambda,\bar\lambda}$, and
$C_{\lambda,\bar\lambda}$ in the matrix forms.  First, we
denote $\hat A$ as the contribution to $A$ at the tree level
in the Standard Model, and $\delta A$ as the deviation due to the CP
violating form factors.
$$
A^V_{\lambda,\bar\lambda}=\hat A_{\lambda,\bar\lambda}
                 + \delta A^V_{\lambda,\bar\lambda} + \cdots
\quad(V=A,Z)
.
$$
The corrections due to other CP conserving
form factors are hidden in the dots. In this paper, we are not
interested in them.
In the basis $(-,0,+)$, the matrices are,
\begin{equation}
\hat A =        \left( \begin{array}{ccccc}
           1     &\quad&    2\gamma            &\quad&     0   \\
          2\gamma&\quad&  1+2\gamma^2          &\quad&     2\gamma\\
             0   &\quad&    2\gamma            &\quad&     1
                     \end{array}  \right)
\;,
\end{equation}
\begin{equation}
\delta A^V  =\left( \begin{array}{ccccc}
-i(\beta^{-1}f_6^V + 4\gamma^2\beta f_7^V)  &\quad&
-i\gamma( f_4^V + \beta^{-1}f_6^V )         &\quad& 0          \\
-i\gamma(-f_4^V + \beta^{-1}f_6^V )         &\quad& 0 &\quad&
 i\gamma( f_4^V + \beta^{-1}f_6^V )                            \\
                                                    0 &\quad&
 i\gamma(-f_4^V + \beta^{-1}f_6^V )         &\quad&
 i(\beta^{-1}f_6^V + 4\gamma^2\beta f_7^V)
\end{array}
\label{eq:cpodd}
\right)
\;.
\end{equation}
Under CP transformation\cite{r:modification}, 
\begin{equation}
   {\cal M}_{\sigma,\bar \sigma;\lambda,\bar \lambda} (\Theta)
\rightarrow
   {\cal M}_{-\bar \sigma, -\sigma; -\bar \lambda,\ -\lambda} (\Theta),
\label{eq:cp-m}
\end{equation}
and therefore
$A_{\lambda,\bar\lambda} \to A_{-\bar\lambda,-\lambda}$, or 
$A_{i,j} \to A_{j,i}$ in present notation.
Therefore form factors $f_{4,6,7}$, as appeared in Eq.(\ref{eq:cpodd}), 
already parametrize the most general CP--odd part of $A$. 
Note also that, under CP$\hat{\hbox{T}}$, 
$A_{\lambda,\bar\lambda} \to A_{-\bar\lambda,-\lambda}^*$, 
therefore the real parts of $f_{4,6,7}$ are 
CP$\hat{\hbox{T}}$--even, but their imaginary parts are 
CP$\hat{\hbox{T}}$--odd.
 
The matrix $B$ is just given by $\hat A$ with the factor 1 at the 00 entry
removed, {\it i.e.}
$B_{\lambda,\bar \lambda}=\hat A_{\lambda,\bar \lambda}
-\delta_{\lambda,0}\delta_{\bar\lambda,0}$.
The  matrix  $C$ is given below,
\begin{equation}
C  =        \left( \begin{array}{ccccc}
\gamma^{-2}&\quad&    (2-2\beta)/\gamma        &\quad& 2\sqrt{2}\beta\\
(2+2\beta)/\gamma&\quad&  2/\gamma^2           &\quad& (2-2\beta)/\gamma\\
2\sqrt{2}\beta &\quad&    (2+2\beta)/\gamma    &\quad& \gamma^{-2}
                     \end{array}  \right)
\;.
\end{equation}

The CP violation in the three gauge boson vertex affects terms with 
$\lambda+\bar\lambda \neq 0$ only.  There are 6 complex phenomenological
CP violating parameters, namely the three form factors, $f_4$, $f_6$,
and $f_7$, each for photon and for $Z$ couplings.  

As stated earlier, the CP information is carried by the helicities of the 
$W$'s.  They can be decoded from the decay products of the $W$ bosons.  
Assuming the standard V$-$A interaction in the $W$ decay, the decay amplitudes 
are 
\begin{eqnarray}
{\cal M}^-( W^- (\lambda) \to {\mbox{f}}_1\bar{\mbox{f}}_2) &\sim& 
l^-_{\lambda} 
= d^1_{\lambda,-1}(\psi)e^{i\lambda\phi},              \nonumber\\
{\cal M}^+( W^+ (\bar \lambda) \to {\mbox{f}}_3\bar{\mbox{f}}_4) &\sim& 
l^+_{\bar\lambda} 
= d^1_{-\bar\lambda,+1}(\bar\psi)e^{-i\bar \lambda\bar\phi}.
\label{eq:ded}
\end{eqnarray}
The normalization is not important in our following discussion.
The polar angle $\psi$ and the azimuthal angle $\phi$ of ${\mbox{f}}_1$
are defined in the $W^-$ rest frame.
Similarly, the polar angle $\bar\psi$ and the azimuthal angle $\bar\phi$
of $\bar{\mbox{f}}_4$ are defined in the $W^+$ rest frame. These two rest
frames of $W^\pm$ are constructed by merely boosting (without rotation)
the $e^-e^+$ CM frame along the common $z$ axis,
which is defined here as pointing in the direction of motion of $W^-$.
In our convention,
there is a sign difference in ${\cal M}^+$ when $\bar\lambda=0$ from that in
Eq.(4.8b) of Ref.\cite{hagiwara}.  (This is consistent with the removal of the 
sign factor $\varepsilon$ from Eq.(\ref{eq:prod}) mentioned earlier).
 
These decay amplitudes have to be folded with the production amplitude
in Eq.(\ref{eq:prod}) to obtain the amplitude for the overall process,
$e^-e^+\to W^-W^+$ followed by $W^-\to {\mbox{f}}_1\bar{\mbox{f}}_2$ and
$W^+\to{\mbox{f}}_3\bar{\mbox{f}}_4$. 
Following Ref.\cite{hagiwara}, the differential 
cross section averaged on  the initial 
fermion polarizations and summed over the final state fermion 
polarizations
can be written as 
\begin{equation}
\sigma_0 P^{\lambda,\bar\lambda}_{\lambda',\bar\lambda'}
(l^-_{\lambda}     l^{-*}_{\lambda'}    )
(l^+_{\bar\lambda} l^{+*}_{\bar\lambda'}) ,
\label{eq:crosssection}
\end{equation} 
where $\sigma_0$ is a factor independent of the $W$ polarizations which 
does not concern us here.  The matrix $P$ is the general 
density matrix of $W^-W^+$ boson pair defined as 
\begin{equation}
P^{\lambda,\bar\lambda}_{\lambda',\bar\lambda'} 
= {\cal N}^{-1} \sum_{\sigma,\bar \sigma}
{\cal M}_{\sigma,\bar \sigma;\lambda,\bar \lambda} (\Theta)
{\cal M}_{\sigma,\bar \sigma;\lambda',\bar \lambda'}^* (\Theta)   .
\end{equation} 
Here ${\cal N}$ is the normalization such that $\hbox{Tr} P$=1.
The azimuthal-angle dependence of 
Eq.(\ref{eq:crosssection}) has been worked out in Ref.\cite{hagiwara}.
Under CP transformation 
\begin{equation}
P^{\lambda,\bar\lambda}_{\lambda',\bar\lambda'} \to 
P^{-\bar\lambda,-\lambda}_{-\bar\lambda',-\lambda'} ,
\end{equation} 
while under CP$\hat{\hbox{T}}$
\begin{equation}
P^{\lambda,\bar\lambda}_{\lambda',\bar\lambda'} \to 
P^{-\bar\lambda',-\lambda'}_{-\bar\lambda,-\lambda} 
= (P^{-\bar\lambda,-\lambda}_{-\bar\lambda',-\lambda'})^*  .
\label{eq:pcp}
\end{equation} 
 
It seems straightforward to measure the asymmetry in event rates between
the two CP conjugated configurations,
\begin{equation}
\hbox{CP:}      (\Theta;\psi,\phi;\bar\psi,\bar\phi)
\leftrightarrow (\Theta;\pi-\bar\psi,\bar\phi+\pi;\pi-\psi,\phi+\pi) \;.
\label{eq:CPconj}
\end{equation}
However, these angle variables may not be completely reconstructed
because (i) the identity of the quark cannot be fully retained in the
final jet configuration, or (ii) the missing neutrinos introduce
ambiguity in event reconstruction.  
 
In addition, in its totally differential form in Eq.(\ref{eq:CPconj}), 
it is not easy to make any simple physical interpretation.
The fact that events scatter over a multivariable domain also
lowers statistics which makes it harder to establish the CP asymmetry.
In the following, we shall try to construct a few CP--odd observables 
which are more accessible to measurement.  
For the decay modes in 
which event reconstruction is possible
(in particular, the mixed lepton--hadron modes), 
we focus on integrated CP asymmetries 
that have simple and intuitive interpretations kinematically.  
For the modes in which only a partial reconstruction is possible, 
such as the purely hadronic modes or the purely leptonic modes, 
we shall construct CP--odd observables based on partial information.
 
Eq.(\ref{eq:CPconj}) provides another way of looking at the 
final state (CP$\hat{\hbox{T}}$-odd) effect.  
It is well-known that the observation of CP violation requires a 
source of complex phase in addition to the one that is due to CP violation.  
Observation of Eq.(\ref{eq:CPconj}) implies that it is possible 
to dig out a CP violating 
signal from either the $\psi$ dependence or from the $\phi$ dependence.  
The complex $\phi$ dependence in Eq.(\ref{eq:ded}) provides automatically 
the additional source of complex phase one needs.  The resulting CP--odd 
observables are therefore can be made CP$\hat{\hbox{T}}$\ even.  They will be  
investigated in Section IV and later.  
To decode the signal from the $\psi$ dependence alone, one needs the form 
factors to be complex to provide the additional 
source of complex phase needed since the the $\psi$-dependent 
$d-$functions in Eq.(\ref{eq:ded}) are real.  
Therefore,  corresponding CP--odd 
observables are CP$\hat{\hbox{T}}$-odd.
 
\section{Leptonic Energy Asymmetry 
(CP--odd and CP$\hat{\hbox{T}}$--odd)}
 
%
In this section we shall start with an CP--odd observable that is 
CP$\hat{\hbox{T}}$--even and therefore requires FSI.  In our case, it means 
only the imaginary parts of the form factors contribute.  
For a process in which the initial state is CP self conjugate and the final 
states are heavy particles with spin, one can easily form a CP--odd quantity 
using different helicity states of the final particles.  
This idea was applied recently to many examples such  
$gg \rightarrow t\bar t$\cite{peskin}, $e^-e^+ \rightarrow t\bar t$
\cite{r:eett}, 
or Higgs decay to $t\bar t$ or gauge boson pairs\cite{r:cphiggs,r:soni}.
To observe this helicity asymmetry one has to rely on the kinematics of the 
heavy particle decay to decode the helicities of the decaying particles.  
Luckily for the top quark and the $W^\pm$ boson this can be done by the 
asymmetry in the energy spectrum of the charged leptons in their semileptonic 
and leptonic decay respectively.  We shall apply this idea here to 
$e^-e^+ \rightarrow W^- W^+$.  
Note that this helicity asymmetry in heavy particle production is 
C--odd( for charged heavy particle pairs),
CP--odd, and CP$\hat{\hbox{T}}$-odd.  Therefore FSI are required to get a 
nonvanishing result and one can only test the imaginary parts of the form 
factors this way.  We shall keep the real parts of the form factors 
at their tree level Standard Model value in this section.
 
Let us begin with a simple illustration of CP violation due to the
absorptive parts of $f_4$, $f_6$ and $f_7$. For the $e^-_R e^+_L$ initial
configuration, only coefficients $A$ contribute to the production
amplitudes in Eq.(\ref{eq:prod}). It is straightforward to see that
the amplitudes ${\cal M}^V_{+,-;\lambda,\bar\lambda}$ due to
$V$ contribution in the $s$--channel are proportional
to coefficients as follows,
\begin{eqnarray}
\tilde{\cal M}^V_{+,-;+,0}
\sim 2+\hbox{Im }f^V_4 -\hbox{Im }f^V_6/\beta ,\nonumber\\
\tilde{\cal M}^V_{+,-;0,-}
\sim 2-\hbox{Im }f^V_4 +\hbox{Im }f^V_6/\beta ,\nonumber\\
\tilde{\cal M}^V_{+,-;0,+}
\sim 2-\hbox{Im }f^V_4 -\hbox{Im }f^V_6/\beta ,\nonumber\\
\tilde{\cal M}^V_{+,-;-,0}
\sim 2+\hbox{Im }f^V_4 +\hbox{Im }f^V_6/\beta          ,
\label{eq:ratediff}\\
\tilde{\cal M}^V_{+,-;+,+}
\sim 1/\gamma - \hbox{Im }f^V_6/(\beta\gamma)
                         -4\gamma\beta\hbox{Im }f_7            ,\nonumber\\
\tilde{\cal M}^V_{+,-;-,-}
\sim 1/\gamma + \hbox{Im }f^V_6/(\beta\gamma)
                         +4\gamma\beta\hbox{Im }f_7            .\nonumber
\end{eqnarray}
The presence of the absorptive parts of $f^V_{4,6,7}$ produces
asymmetry in the production rates between the CP conjugate states
$(\lambda,\bar\lambda)$ and $(-\bar\lambda,-\lambda)$, {\it i.e.}
$(+,0)$ and $(0,-)$; $(0,+)$ and $(-,0)$; or $(+,+)$ and $(-,-)$.
Such CP asymmetry also happens in the $e^-_Le^+_R$ channel
although the formulas become much more lengthy because of the involvement
of the $\nu$--exchange diagram.  However, it is just as straightforward 
to study the amplitudes in Eqs.(\ref{eq:prod},\ref{eq:amplitude}) numerically.
This is what we shall do here, instead of 
producing the complete analytic formula.   
 
Since we are going to integrate over the azimuthal angles $\phi$ and 
$\bar\phi$, and assuming that the $W$ bosons are produced on shell, 
the resulting $W^-W^+$ production cross section is proportional to 
\begin{equation}
\sigma_{\lambda,\bar\lambda} = 
{\cal N} \sum_{\lambda',\bar\lambda'}
P^{\lambda,\bar\lambda}_{\lambda',\bar\lambda'}
\delta_{\lambda,\lambda'}\delta_{\bar\lambda,\bar\lambda'}
= \sum_{\sigma,\bar \sigma}
{\cal M}_{\sigma,\bar \sigma;\lambda,\bar \lambda} (\Theta)
{\cal M}_{\sigma,\bar \sigma;\lambda,\bar \lambda}^* (\Theta)  .
\label{eq:pmatrix}
\end{equation} 
The interference effect between different
($\lambda$,$\bar\lambda$) configurations drops away 
because the integration over the azimuthal
angles kills the ``off--diagonal'' contributions.
There are three CP violating rate differences, namely,
$\sigma_{+,0}-\sigma_{0,-}$, $\sigma_{0,+}-\sigma_{-,0}$,
and $\sigma_{+,+}-\sigma_{-,-}$. 
The detailed relationship between $\sigma_{\lambda,\bar\lambda}$ and 
the form factors is very tedious and 
has been worked out by Gounaris et. al. in 
Ref.\cite{gounaris2}.  Here we shall concentrate on the contributions 
$\sigma_{\lambda,\bar\lambda}$ to the lepton energy asymmetry.
However one should note that in general 
$\sigma_{+,0}-\sigma_{0,-}$ depends only on the combination 
$\hbox{Im }f^V_4 -\hbox{Im }f^V_6/\beta$ and $\sigma_{0,+}-\sigma_{-,0}$ 
only on $\hbox{Im }f^V_4 +\hbox{Im }f^V_6/\beta$
and $\sigma_{+,+}-\sigma_{-,-}$ only on 
$\hbox{Im }f^V_6/(\beta\gamma)+4\gamma\beta\hbox{Im }f_7$.

Consider events with one of the $W$ bosons decaying leptonically. The
other $W$ can decay either hadronically or leptonically. The energy
spectra of the lepton $l$ from $W^-$ and the anti--lepton $\bar l$ from
$W^+$ will be different due to the above asymmetry.
In the CM frame of the collider, the energies of the lepton $l$ and the
anti--lepton $\bar l$ are determined by the variables
$\psi$ and $\bar \psi$ respectively.
\begin{eqnarray}
E(\ell^-) & = & \case 1/4 \sqrt{s} (1+\beta \cos\psi) \;,\nonumber\\
E(\ell^+) & = & \case 1/4 \sqrt{s} (1-\beta \cos\bar \psi) .
\end{eqnarray}
Define the energy fraction $x=4E/\sqrt{s} \in [1-\beta,1+\beta]$.
The lepton (anti--lepton) from the decay of a right handed
$W^-_{\lambda=1}$ (left handed $W^+_{\bar\lambda=-1}$) will
have a softer energy  spectrum,
\begin{equation}
  f(x) = \case 3/8 \beta^{-3} (x-1-\beta)^2  .
\end{equation}
This normalized function is derived based on Eq.(\ref{eq:ded}).
Similarly, the lepton (anti--lepton) from the decay of a left handed
$W^-_{\lambda=-1}$ (right handed $W^+_{\bar\lambda=1}$)
will have a harder energy spectrum,
\begin{equation}
  g(x) = \case3/8 \beta^{-3} (x-1+\beta)^2  .
\end{equation}
And, the lepton (anti--lepton) from the decay of a longitudinal
$W^-_{\lambda=0}$ ($W^+_{\bar\lambda=0}$) will have
an energy spectrum symmetric with respect to the average value $x=1$,
\begin{equation}
n(x) = \case3/4 \beta^{-3} (\beta^2-1+2x-x^2) 
= \case3/2 \beta^{-1} - f(x) - g(x) .
\end{equation}
We obtain the energy asymmetry
\begin{eqnarray}
a_E(x) \equiv {1\over N^-}{dN^-\over dx(\ell^-)} 
- {1\over N^+}{dN^+\over dx(\ell^+)}
=\Bigl( \sum_{\lambda,\bar \lambda} \sigma_{\lambda,\bar\lambda}
 \Bigr)^{-1} \Bigl\{
     (\sigma_{+,+}-\sigma_{-,-})[f(x)-g(x)]              \nonumber\\
   + (\sigma_{+,0}-\sigma_{0,-})[f(x)-n(x)]
   - (\sigma_{0,+}-\sigma_{-,0})[g(x)-n(x)]              
\Bigr\} \;                                               \nonumber\\
=\Bigl( \sum_{\lambda,\bar \lambda} \sigma_{\lambda,\bar\lambda}
 \Bigr)^{-1} 
\Bigl\{ \Bigl(
\sigma_{+,0}-\sigma_{0,-} +\sigma_{0,+}-\sigma_{-,0}
           +2\sigma_{+,+}-2\sigma_{-,-}
\Bigr)
 \case1/2[f(x) -g(x)]                                    \nonumber\\
 + 
\Bigl(
\sigma_{+,0} - \sigma_{0,-} - \sigma_{0,+} + \sigma_{-,0}
\Bigr)
 \case3/2 [f(x) + g(x) - \beta^{-1}]
\Bigr\} \;.
\label{eq:energyasymm}
\end{eqnarray}
Here distributions are compared at the same energy for the lepton and the
anti--lepton, $x=x(\ell^+)=x(\ell^-)$. The count $N^-$ ($N^+$) includes events
with a prompt lepton (anti--lepton) from $W^-$ ($W^+$) decay 
with $W^+$ ($W^-$) decays arbitrarily.
Here the cross sections $\sigma_{\lambda,\bar\lambda}$ are still in general 
a function of the variable $\Theta$. 
%
This equation also implies that there are only two CP--odd combinations of 
$\sigma_{\lambda,\bar\lambda}$ one can probe.  One of them is P-even. 
The other one is P-odd.  To improve the observability, one may as well 
integrate $a_E(x)$ over ranges of $x$.
It is also obvious that the asymmetry will cancel if we integrate
$a_E(x)$ over the complete range of $x$, namely from $1-\beta$ to
$1+\beta$. One easy way to define the integrated energy asymmetry, $A_E$
is to integrate $a_E(x)$ over the upper half range, {\it i.e.} from 1 to
$1+\beta$,
\begin{eqnarray}
A_E &\equiv& \left(\int_1^{1+\beta}
                -\int_{1-\beta}^1\right) a_E(x)dx  \nonumber\\
&=& -\case3/4
\Bigl(
\sigma_{+,0}-\sigma_{0,-} +\sigma_{0,+}-\sigma_{-,0}
           +2\sigma_{+,+}-2\sigma_{-,-}
\Bigr) /
\Bigl(
   \sum_{\lambda,\bar \lambda} \sigma_{\lambda,\bar\lambda}
              \Bigr)
\;.
\end{eqnarray}
Assuming that the CP odd form factors are small perturbation from the
CP conserving ones, we can compute the expected energy
asymmetry per unit of Im~$f_i$ as a function of $\cos\Theta$.  
This is plotted in Fig.~2a,~2b for different center-of-mass energy.  
The asymmetry $A_E$ is smaller in the forward region, $\cos\Theta \sim 1$, 
because in that region the CP conserving neutrino mediated diagrams dominate.  
In the backward region, $\cos\Theta \sim -1$, both Im~$f_4^{\gamma,Z}$ give 
positive contributions while Im~$f_6^{\gamma,Z}$ and Im~$f_7^{\gamma,Z}$ 
give negative contributions.  

Also note that $A_E$ is a parity-odd (C-even) operator.  Therefore, if one 
considers the S-channel photon-mediated diagrams alone, the form factor 
Im~$f_4^{\gamma}$ will not contribute.  However, since the lowest order 
diagrams also include the $Z$-mediated graphs which have a different parity 
property, Im~$f_4^{\gamma}$ still contributes even if 
the lowest order neutrino mediated diagram is negligible.
 
The integration limits are arbitrary.  If one integrates over
$x$ symmetrically about the midpoint 1, then the contribution from
$\sigma_{+,+}-\sigma_{-,-}$ will be eliminated because $f(x) - g(x)$ 
is antisymmetric about $x=1$.  Therefore there is no contribution from 
Im~$f_7$ in this case.  These types of combinations are parity-even and C-odd.  
However, just as for Im~$f_4^{\gamma}$ in the previous case, 
Im~$f_6^{\gamma}$ also contributes even if the neutrino 
dominated diagrams are negligible.  
For example, one can define
\begin{eqnarray}
A'_E(\alpha) &\equiv& \left(
 \int_{1-\alpha}^{1+\alpha}
-\int_{1-\beta}^{1-\alpha}
-\int_{1+\alpha}^{1+\beta}
            \right)
a_E(x)dx                   \nonumber\\
&=& -n_\alpha
\Bigl(
\sigma_{+,0}-\sigma_{0,-}-\sigma_{0,+}+\sigma_{-,0}
\Bigr)/
\Bigl(
\sum_{\lambda,\bar \lambda} \sigma_{\lambda,\bar\lambda}
\Bigr)
\;,
\end{eqnarray}
where $0 < \alpha < \beta$ and $n_\alpha$ is a numerical constant.   
For $\alpha = \beta/3$, $n_\alpha = \case{12}/{27}$.
The contribution of various Im~$f_i$ to $A'_E$ is plotted in Fig.2c 
as a function of $\cos\Theta$.
Just as in the case of asymmetry $A_E$, $A_E'$ is smaller in the forward 
region because of the neutrino mediated diagrams.  
In the backward region, the contributions of both Im~$f_4^{\gamma,Z}$ 
are positive while those of Im~$f_6^{\gamma,Z}$ are negative.  

From Eq.(\ref{eq:energyasymm}), one can not probe all three independent 
CP--odd combinations of $\sigma_{\lambda,\bar\lambda}$.  
Therefore the lepton energy asymmetry alone can not probe all three 
independent imaginary parts of the form factors, Im~$f_{4,6,7}$.  
We shall discuss another probe of the these imaginary parts later.
In addition, two choices of the integrated asymmetry presented above 
are just examples which may not be optimal.
A careful weighting of events to avoid cancellations
may still significantly improve the sensitivity of the asymmetry.

In Eq.(\ref{eq:energyasymm}) and in the figures, 
we have summed over the total decay modes of the other $W$ boson.  
However, there may be other contributions to the leptonic energy asymmetry 
not going through the two $W$ intermediate states.  If one does not use at 
least one of the $W$ bosons as a tag to eliminate other contributions, one 
will still have to investigate the other potential contributions before the 
measurement can be interpreted.  However, one can also 
tag the other $W$ boson by selecting only 
its purely hadronic decay modes and use the jet energy and 
momentum measurement to eliminate the other potential contributions which 
are not due to $W$ pair production.  Of course this will require a 
sacrifice in event rate.  The detection uncertainty in jet energy and 
momentum also has to be taken into account.  However, even with this 
uncertainty, the efficiency in eliminating the non-$WW$ background can be 
still quite high.

\section{Uncorrelated up--down asymmetry.}
\label{sec:uncorrelated}
 
It is known that explicit CP violation requires the CP nonconserving
vertex as well as additional complex amplitudes. In the previous case, this
complex structure comes from the absorptive part due to the final state
interactions. However, the complex structure can also come from other
sources. One of these is the azimuthal phase $\exp(iL_z\phi)$ in the
decay process. This will produce a CP--odd up--down
asymmetry even with only the dispersive part of CP violating vertex,
Re $f_{4,6,7}$.
 
We concentrate our attention to those events that one of the $W$
decays leptonically and the other one decays into quark jets, {\it i.e.}
either ${\mbox{f}}_1=\ell^-$ or $\bar{\mbox{f}}_4=\ell^+$. 
CP violation could appear
as the difference of two separate azimuthal distributions for $\ell^-$ and
$\ell^+$.
This kind of angular asymmetry is very simple. We only look at the
azimuthal angle of the lepton or the anti--lepton from one $W$. The
recoiling jets from the other $W$ are only used to define the reaction
plane.
In this way, we are looking at the separate (uncorrelated)
density matrices for $W^-$ and $W^+$.
 
The angular distribution of $\ell^-$ from the $W^-$ decay is specified by the
the spin density matrix $\rho_{\lambda,\lambda'}$ of the $W$ boson.
\begin{equation}
   \rho(\Theta)_{\lambda,\lambda'}
      ={\cal N}(\Theta)^{-1}\sum_{\sigma,\bar\sigma,\bar\lambda}
  {\cal M}  (\sigma,\bar\sigma,\lambda ,\bar\lambda)
  {\cal M}^*(\sigma,\bar\sigma,\lambda',\bar\lambda)
      =\sum_{\bar\lambda}P^{\lambda,\bar\lambda}_{\lambda',\bar\lambda'}
\;.
\end{equation}
Here ${\cal N}$ is the normalization such that $\hbox{Tr}\rho$=1.
$\rho$ is hermitian by definition. The normalized distribution
for $\ell^-$ is given by
\begin{eqnarray}
   {dN(\ell^-)\over d\phi d\cos\psi}={1\over 4\pi}{3\over4}
             \Bigl[(1-\cos\psi)^2\rho_{++}
                + (1+\cos\psi)^2\rho_{--}
                +2\rho_{00}  \sin^2\psi \nonumber\\
    -2\sqrt{2} \hbox{Re }\rho_{+,0}(1-\cos\psi)\sin\psi\cos\phi
    +2\sqrt{2} \hbox{Im }\rho_{+,0}(1-\cos\psi)\sin\psi\sin\phi
                                              \nonumber\\
    -2\sqrt{2} \hbox{Re }\rho_{-,0}(1+\cos\psi)\sin\psi\cos\phi
    -2\sqrt{2} \hbox{Im }\rho_{-,0}(1+\cos\psi)\sin\psi\sin\phi
                                              \nonumber\\
    +2 \hbox{Re }\rho_{+,-}(1-\cos^2\psi)\cos2\phi
    -2 \hbox{Im }\rho_{+,-}(1-\cos^2\psi)\sin2\phi \Bigr]
    \;.
\label{eq:phi}
\end{eqnarray}
Similarly, the angular distribution of $\ell^+$ from the $W^+$ decay is
specified by another spin density matrix,
$\bar\rho_{\bar \lambda,\bar\lambda'}$, of the $W^+$ boson.
\begin{equation}
   \bar \rho(\Theta)_{\bar \lambda,\bar \lambda'}
      ={\cal N}(\Theta)^{-1} \sum_{\sigma,\bar\sigma,\lambda}
  {\cal M}  (\sigma,\bar\sigma,\lambda,\bar \lambda)
  {\cal M}^*(\sigma,\bar\sigma,\lambda,\bar \lambda')\;.
\end{equation}
The nomalized distribution for $\ell^+$ is given by
\begin{eqnarray}
   {dN(\ell^+)\over d\bar\phi d\cos\bar\psi}={1\over 4\pi}{3\over4}
    \Bigl[(1-\cos\bar\psi)^2\bar\rho_{++}
                + (1+\cos\bar\psi)^2\bar\rho_{--}
                +2\bar\rho_{00}\sin^2\bar\psi \nonumber\\
    +2\sqrt{2} \hbox{Re }\bar\rho_{+,0}(1-\cos\bar\psi)
                  \sin\bar\psi\cos\bar\phi
    +2\sqrt{2} \hbox{Im }\bar\rho_{+,0}(1-\cos\bar\psi)
                  \sin\bar\psi\sin\bar\phi
                                              \nonumber\\
    +2\sqrt{2} \hbox{Re }\bar\rho_{-,0}(1+\cos\bar\psi)
                  \sin\bar\psi\cos\bar\phi
    -2\sqrt{2} \hbox{Im }\bar\rho_{-,0}(1+\cos\bar\psi)
                  \sin\bar\psi\sin\bar\phi
                                              \nonumber\\
    +2 \hbox{Re }\bar\rho_{+,-}(1-\cos^2\bar\psi)\cos2\bar\phi
    +2 \hbox{Im }\bar\rho_{+,-}(1-\cos^2\bar\psi)\sin2\bar\phi
    \Bigr]  \;.
\label{eq:phibar}
\end{eqnarray}
In our present phase convention,
if CP were conserved, we would have
the following identities, first noticed by
Gounaris {\it et al.}\cite{gounaris,r:modification},
\begin{equation}
\rho(\Theta)_{\lambda,\lambda'}=\bar\rho(\Theta)_{-\lambda,-\lambda'}
\; .
\label{eq:Gou}
\end{equation}
Under CP conjugation, we exchange $\ell^-$ and $\ell^+$
with angle substitutions $\bar\psi \leftrightarrow \pi-\psi$,
$\bar\phi \leftrightarrow \pi+\phi$.
The distribution in Eq.(\ref{eq:phibar})
is transformed into Eq.(\ref{eq:phi}) provided Eq.(\ref{eq:Gou}) is
satisfied.
On the other hand, the CP$\hat{\hbox{T}}$\ invariance
implies\cite{hagiwara,gounaris}.
\begin{equation}
\rho(\Theta)_{\lambda,\lambda'}=\bar\rho(\Theta)^*_{-\lambda,-\lambda'}
\; .
\label{eq:GouCPT}
\end{equation}
When the effect of the CP violating form factors are included in the
analysis, one can form the following CP or CP$\hat{\hbox{T}}$\ 
eigen--combinations:
$$R(\pm)(\Theta)_{\lambda,\lambda'} =
\hbox{Re } \rho(\Theta)_{\lambda,\lambda'}
\pm \hbox{Re } \bar\rho(\Theta)_{-\lambda,-\lambda'}   \;,$$
and
$$I(\pm)(\Theta)_{\lambda,\lambda'} =
\hbox{Im } \rho(\Theta)_{\lambda,\lambda'}\pm
\hbox{Im } \bar\rho(\Theta)_{-\lambda,-\lambda'}       \;.$$
Among them, $R(-)$ and $I(+)$ are CP$\hat{\hbox{T}}$\ odd;
$R(-)$ and $I(-)$ are CP odd and the others are either 
CP$\hat{\hbox{T}}$- or CP--even respectively.
The observation of CP--odd $R(-)$ requires final state interactions
due to CP$\hat{\hbox{T}}$.  Therefore it does not need to involve 
the complex
phase of the azimuthal dependence in Eqs.(\ref{eq:phi}, \ref{eq:phibar}).
It can be decoded by analyzing the polar angular dependence
of $\psi$ or $\bar\psi$ in these equations.
In the collider M frame these
dependences can be translated into the
energy dependence of the corresponding lepton in the final state,
which has already been studied in the previous section.

Here we shall focus on $I(-)$ which does not require
final state interactions.
Since $\rho$ is hermitian, the only nonzero
components of $I(-)$ is $I(-)_{+,-}$, $I(-)_{+,0}$, $I(-)_{0,-}$.
They are related to Re $f_{4,6,7}$.

One can in principle make detailed angular analysis of the difference
between Eq.(\ref{eq:phi}) and its CP conjugate in Eq.(\ref{eq:phibar})
similar to what was done for the case of $e^-e^+\rightarrow W^-W^+$
by Gounaris {\it et al.}\cite{gounaris}
However, since all experimental measurements of angles have finite 
resolution, such analysis is not very useful in practice.  
We wish to find simpler observables which may be more
intuitive and may be easier to measure, we shall consider the following
partially integrated observable.
Let $dN(\ell^+,\hbox{up})$ count events with $\ell^+$ from $W^+$ decay 
emitted above the $xz$ plane, {\it i.e.} $p_y(\ell^+)>0$ where $xz$ plane 
is defined by the $q_1\bar{q}_2$ pair of the $W^-$ decay; and 
$dN(\ell^-,\hbox{up})$ similarly.  
Then, with other obvious notations, we define the following up--down
asymmetry 
\begin{equation}
{\cal A}^{u.d.}(\Theta) =
{[dN(\ell^-,\hbox{up})+dN(\ell^+,\hbox{up})]
-[dN(\ell^-,\hbox{down})+dN(\ell^+,\hbox{down})]
\over
[dN(\ell^-,\hbox{up})+dN(\ell^+,\hbox{up})]
+[dN(\ell^-,\hbox{down})+dN(\ell^+,\hbox{down})]
}\;.
\label{eq:Aud}
\end{equation}
It is evaluated for each scattering angle $\Theta$. The branching
fraction of the $W$ semileptonic decay cancels in the ratio. Integrating
on $\psi$ and $\bar\psi$ over the full range and on $\phi, 
\bar\phi$ over up or down hemispheres,
we obtain ${\cal A}^{u.d.}(\Theta)$ in a very simple form from
Eqs.(\ref{eq:phi},\ref{eq:phibar}),
\begin{eqnarray}
{\cal A}^{u.d.}(\Theta)&=&\case3/8\sqrt{2}\Bigl[
     \hbox{Im } \rho(\Theta)_{+,0}
    -\hbox{Im } \bar\rho(\Theta)_{-,0}
    -\hbox{Im } \rho(\Theta)_{-,0}
    +\hbox{Im } \bar\rho(\Theta)_{+,0} \Bigr]  \nonumber \\
	&=&\case3/8\sqrt{2}\Bigl[ I(-)_{+,0} - I(-)_{-,0} \Bigr].
\label{eq:Ad}
\end{eqnarray}
As we sum up contributions from $\ell^\pm$ in each square bracket of
Eq.(\ref{eq:Aud}),
the asymmetry is insensitive to the sign of charge,
it is obvious that a non-vanishing value
of ${\cal A}^{u.d}(\Theta)$ is a genuine signal
of CP violation. Though, the angular distributions of the leptons 
derived in Eqs.(\ref{eq:phi},\ref{eq:phibar}) 
will have corrections from other CP conserving final state
interactions, 
however, these corrections cannot fake the CP asymmetry as
the effects due to the CP conserving sources cancel away in the differences.

To make distinction between $I(-)_{+,0}$ and $I(-)_{-,0}$ one can not 
integrate over the full range of $\psi$ or $\bar\psi$.
If we restrict $dN(\ell^\pm)$ in the numerator of Eq.(\ref{eq:Aud}) to
count only events with 
$E(\ell^\pm) > E_0$ for some energy $E_0$ in the physical 
range, $\case1/4\sqrt{s}(1 + \beta) \ge E_0 \ge \case1/4\sqrt{s}(1 - \beta)$,
and keep the denominator
unchanged, then one obtain a different
integrated asymmetry ${\cal A'}^{u.d.}$.   One can show that
\begin{eqnarray}
{\cal A'}^{u.d.}(\Theta)={3\sqrt{2}\over4\pi} & \Biggl\{&
  \bigl(
a(E_0) - b(E_0)
\bigr)
  \bigl(\hbox{Im }     \rho(\Theta)_{+,0}
       -\hbox{Im } \bar\rho(\Theta)_{-,0}
  \bigr)
 \nonumber\\
 &-& \bigl(
a(E_0) + b(E_0)
\bigr)
   \bigl( \hbox{Im }     \rho(\Theta)_{-,0}
         -\hbox{Im } \bar\rho(\Theta)_{+,0}
   \bigr)                                     \Biggr\},
\label{eq:Apud}
\end{eqnarray}
which gives a different combination of $I(-)_{+,0}$ and $I(-)_{-,0}$.
Here, $a(E_0) = - \case1/2\sin\eta\cos\eta + \case1/2\eta$, and 
$b(E_0) = \case1/3\sin^3\eta$ with
$E_0 = \case1/4\sqrt{s}(\beta\cos\eta +1)$.  For example, if we use 
$E_0 = \case1/4\sqrt{s}$, the average energy, then $\eta = \pi/2$ 
and $a(E_0) = \pi/4$, $b(E_0) = 1/3$.

To probe the effect of the CP violating quantity 
$I(-)_{+,-}$ one has to look into the azimuthal 
dependence.  For example, in another integrated asymmetry such as
\begin{equation}
{\cal A''}^{u.d.}(\Theta) =
{
 [dN(\ell^-,\hbox{I+III})+dN(\ell^+,\hbox{II+IV})]
-[dN(\ell^-,\hbox{II+IV})+dN(\ell^+,\hbox{I+III})]
\over
[dN(\ell^-,\hbox{I+II+III+IV})+dN(\ell^+,\hbox{I+II+III+IV})]
}\;.
\label{eq:Appud}
\end{equation}
Here the range of the azimuthal angle has been divided into four usual
quadrants I,II,III and IV.
It can be shown that
\begin{equation}
{\cal A''}^{u.d.}(\Theta) =-{1\over\pi}
   \Bigl(\hbox{Im } \rho(\Theta)_{+,-}
    -\hbox{Im } \bar\rho(\Theta)_{-,+} \Bigr)
\;.
\label{eq:Apd}
\end{equation}
The above three independent measurements are enough to extract the three
CP--odd (CPT--even) parameters $I(-)_{+,-}$, $I(-)_{+,0}$,
$I(-)_{-,0}$ which are related to the elements of the density matrices.

The relationship between $I(-)_{+,-}$, $I(-)_{+,0}$, $I(-)_{-,0}$ 
and the CP violating form factors are too lengthy to be useful to 
present explicitly here.  
In Fig.~3(a,b,c), we plot the contributions to 
$I(-)_{+,0}$, $I(-)_{-,0}$, $I(-)_{+,-}$ due to various form factors.  
Re~$f_7^{\gamma,Z}$ give rise to the largest contributions to 
$I(-)_{+,0}$, especially in the region $\cos\theta < 0$; 
while Re~$f_6^{\gamma,Z}$ give the largest contributions to 
$I(-)_{-,0}$ especially in the region $\cos\theta < 0$.

\section{CP violation in Purely Hadronic $W$ Decays}
\label{sec:hadron}
 
When the $W$ decays into quarks which turn into jets, the information 
about the charges and flavors of the quarks is lost.  The useful 
definitions of the angles depend on the 
amount of information which can be measured.  
In the ideal case, when the quark charge and flavor can be detected, 
then unambiguous definitions of the angles used in previous sections 
can be carried over to the hadronic case 
by substituting a quark doublet for a lepton doublet. 
However, in reality, the flavor and charge can not be identified 
event by event, so none of the observables discussed 
in previous sections can be uniquely defined.
A better starting point is to use the angles that can be defined purely 
kinematically and which do not refer to the flavors or the charges of the 
underlying quarks or $W$ bosons.  
For example, one can define the variables by singling out one of 
the jets kinematically.  
Such an approach has been proposed before for other processes\cite{r:CPjet}.
In this section we wish to present two 
different but similar observables for this purpose.

To start, one can define $W^{(>)}$ to be the $W$ which
has the the hardest jet.  
Since the quark with the largest energy belongs to the $W$ boson 
with the largest $|\cos\psi|$, where $\psi$ of the decay products
is the polar angle at the $W$ rest frame defined after Eq.(11). 
$W^{(>)}$ is either $W^-$ if $|\cos\psi| > |\cos\bar\psi|$, or
$W^+$ if $|\cos\psi| < |\cos\bar\psi|$.
In practice, one has to impose some cuts to exclude events for which
$|\cos\psi| \simeq |\cos\bar\psi|$.  
Once the $W^{(>)}$ is selected one can use it to define an 
unique orientation for the production plane.
One can define uniquely a new scattering angle
$\Theta^{(>)}$, using the $W^{(>)}$ instead of the 
undetermined $W^{-}$.  
 
The C, P, CP, and CP$\hat{\hbox{T}}$\ transformations of a typical
$e^-e^+\rightarrow W^-W^+ \rightarrow q_1 \bar{q_2} q_3 \bar{q_4}
\rightarrow 4$-jet event is 
demonstrated in Fig.~4.  One very simple CP violating observable 
that one can easily identify is 
the forward-backward asymmetry of the most energetic jet.  
The observable is also CP$\hat{\hbox{T}}$--odd, 
therefore can be used to study the imaginary parts of the 
CP violating form factors.  
It is also possible to study the real parts of the form factors, 
however one has to look into the azimuthal dependence of jets.  
First, we look at the asymmetry in the fully differential 
cross section
$\sigma^{(>)}(\Theta^{(>)}; \psi_>,\phi_>;\psi_>,\phi_<)$
where $\psi_>, \phi_>$ are the CM angles of $W^{(>)}$ and 
$\psi_<, \phi_<$ are the corresponding angles of the other $W$.  
One can define the general CP--odd asymmetry $A_{CP}$ as 
\begin{equation}
A_{CP}=\sigma^{(>)}(\Theta^{(>)};\psi_>,\phi_>;\psi_<, \phi_<)
-\sigma^{(>)}(\pi-\Theta^{(>)}; 
\psi_>,\pi-\phi_>;\psi_<,\pi- \phi_<).
\end{equation}
Meanwhile, under CP$\hat{\hbox{T}}$\ 
\begin{equation}
\sigma^{(>)}(\Theta^{(>)}; \psi_>,\phi_>;\psi_<, \phi_<)
\rightarrow
\sigma^{(>)}(\pi-\Theta^{(>)}; \psi_>, \pi + \phi_>;\psi_<, \pi + \phi_<),
\end{equation}
which clearly indicates that if one integrate over full range of 
$\psi_>, \phi_>, \psi_<$ and $\phi_<$ the resulting asymmetry is 
CP$\hat{\hbox{T}}$-odd.
One can also form a more general CP--odd, CP$\hat{\hbox{T}}$-odd 
differential asymmetry to measure 
the absorptive form factors such as 
\begin{eqnarray}
A_i & = & \sigma^{(>)}(\Theta^{(>)}; 
\psi_>,\phi_>;\psi_<, \phi_<)
 -\sigma^{(>)}(\pi-\Theta^{(>)};
\psi_>,\pi-\phi_>;\psi_<,\pi- \phi_<)
\nonumber\\
&-&\sigma^{(>)}(\pi-\Theta^{(>)}; 
\psi_>, \pi + \phi_>;\psi_<, \pi + \phi_<)
+\sigma^{(>)}(\Theta^{(>)};
\psi_>, - \phi_>;\psi_<, - \phi_<).
\end{eqnarray}

Since CP and CP$\hat{\hbox{T}}$\ transformations differ in the
azimuthal distributions one has to dig into this dependence to 
probe the CP--odd, CP$\hat{\hbox{T}}$-even part.  
For example one can form the corresponding 
CP$\hat{\hbox{T}}$-even differential asymmetry 
which measured the real part
of the CP violating form factors:
\begin{eqnarray}
A_r & = & \sigma^{(>)}(\Theta^{(>)}; \psi_>,\phi_>;\psi_<, \phi_<)
 -\sigma^{(>)}(\pi-\Theta^{(>)}; 
\psi_>,\pi-\phi_>;\psi_<,\pi- \phi_<)
\nonumber\\
&+&\sigma^{(>)}(\pi-\Theta^{(>)}; 
\psi_>, \pi + \phi_>;\psi_<, \pi + \phi_<)
-\sigma^{(>)}(\Theta^{(>)};
\psi_>, - \phi_>;\psi_<, - \phi_<).
\end{eqnarray}
 
After integrating decay angles,
the differential cross section $\sigma^{(>)}$ for producing $W^{(>)}$
with the hardest jet at an angle $\Theta^{(>)}$ can
be related to an average of the underlying $W^+ W^-$ production 
cross section.  It can be written as
\begin{eqnarray}
\sigma^{(>)}(\Theta^{(>)}) &=&
 \sum_{\lambda,\bar\lambda}
 \sigma_{\lambda,\bar\lambda}(\Theta^{(>)})
\int_0^1 d\zeta\int_0^1 d\bar\zeta
 D_\lambda(\zeta) D_{\bar\lambda}(\bar\zeta)
 \vartheta(\zeta-\bar\zeta)                      \nonumber\\
&+&\sum_{\lambda,\bar\lambda}
 \sigma_{\lambda,\bar\lambda}(\pi-\Theta^{(>)})
\int_0^1 d\zeta\int_0^1 d\bar\zeta
 D_\lambda(\zeta) D_{\bar\lambda}(\bar\zeta)
 \vartheta(\bar\zeta-\zeta) \;,
\end{eqnarray}
where the first sum is the contribution 
when the hardest jet is originated from $W^-$ 
while the second sum is the contribution 
when the hardest jet is originated from $W^+$.
The decay probability function
$D_\lambda(\zeta)$ is related to those decay amplitudes in Eq.(\ref{eq:ded}),
\begin{equation}
D_\lambda(\zeta)=\case3/2B_h(|d^1_{\lambda,-1}(\psi)|^2
                         +|d^1_{\lambda,-1}(\pi-\psi)|^2)
                =\left\{ \begin{array}{ll}
                  \case3/4B_h (1+\zeta^2)    & \mbox{if $\lambda=\pm1$,}\\
                  \case3/2B_h (1-\zeta^2)    & \mbox{if $\lambda=0$.}
                 \end{array}  \right.
\;,
\end{equation}
where we relate $\zeta=\cos\psi$, etc.
The coefficients, ${3\over2}$ or ${3\over4}$, normalize the functions 
to one after integration over $\zeta$. 
$B_h$ is the branching fraction for the $W$
boson decaying into the purely hadronic mode. We have normalized 
$D_\lambda(\zeta)$ to $B_h$ when it is integrated over $0\le\zeta\le1$.
The step function $\vartheta(\zeta)$ is unity if $\zeta>\epsilon_c$, zero
otherwise.  The threshold $\epsilon_c$ should correspond to the minimum
jet energy difference detectable corresponding to the experimental 
resolution, but for simplicity,
we illustrate the ideal case that $\epsilon_c=0$.

Using the following integrals
\begin{eqnarray}
\int_0^1 d\zeta \int_0^1 d\bar{\zeta}\;
 (1 + s \zeta^2)(1 + s' \bar \zeta^2) \;\vartheta(\zeta-\bar{\zeta})
=  \case1/4[(2 + s) + s' (\case1/3 + s \case2/9)]   \;,
\end{eqnarray}
where $s, s'$ are two sign factors taking values $\pm1$,
The forward--backward asymmetry is proportional to the 
difference of the following two integrals,

\begin{eqnarray}
\int_0^1 d\zeta \int_0^1 d\bar{\zeta}\;
 (1 + s \zeta^2)(1 + s' \bar \zeta^2) \Bigl[ \vartheta(\zeta-\bar{\zeta}) 
- \vartheta(\bar{\zeta}-\zeta) \Bigr]
\nonumber\\ 
%
=\case1/6 (s-s').
\end{eqnarray}
In this equation, the sign $s = \pm 1$ corresponds to $|{\lambda}|=1$ and 
$|{\lambda}|=0$ respectively, and similarly, the sign $s' = \pm 1$
corresponds to either $|\bar\lambda|=1$ and $|\bar\lambda|=0$ respectively.  
Note that the contribution vanishes unless 
$\lambda \bar\lambda = 0$.  Therefore, this asymmetry is not going to give 
us information about the CP--odd $\sigma_{+,+} - \sigma_{-,-}$. 
This is because after one averages over the angles 
$\psi$ and $\pi-\psi$, there is no
difference between the decay of a $W^-$ with helicity $+$ and a $W^-$ with
helicity $-$.  For this reason, 
the asymmetry is insensitive to Im~$f_7$.
\def\dump{
This difference can be written as a matrix $V$,
\begin{equation}
V =\frac{3}{8}
\left( \begin{array}{rcl}
0 \quad&\quad 1 \quad&\quad 0 \\
-1\quad&\quad 0  \quad&\quad -1 \\
0 \quad&\quad 1 \quad&\quad 0
\end{array} \right)  \;,
\end{equation}
where the rows and columns are $\lambda=-1,0,+1$ and $\bar{\lambda}=-1,0,+1$
respectively.}
With this, we can derive the forward-backward asymmetry of the hardest jet
analytically as 
\begin{eqnarray}
 \sigma^{(>)}(\Theta^{(>)})-\sigma^{(>)}(\pi-\Theta^{(>)})
= \case3/8 B_h^2 &\bigl[&
   \sigma_{-,0}(\Theta^{(>)}) - \sigma_{0,+}(\Theta^{(>)})    \nonumber\\
&+&\sigma_{+,0}(\Theta^{(>)}) - \sigma_{0,-}(\Theta^{(>)})    \nonumber\\
&-&   (\Theta^{(>)} \to \pi-\Theta^{(>)})  \bigr]            \;.
\label{eq:asyHJ}
\end{eqnarray}
The asymmetry is usually defined as the ratio of the above quantity to
the symmetric sum as below,
\begin{equation}
 \sigma^{(>)}(\Theta^{(>)})+\sigma^{(>)}(\pi-\Theta^{(>)})
= B_h^2 \sum_{\lambda,\bar \lambda}
[ \sigma_{\lambda,\bar \lambda}(\Theta^{(>)}) 
+ \sigma_{\lambda,\bar \lambda}
(\pi-\Theta^{(>)})].
\end{equation}
The branching fraction will drop away in the ratio. This asymmetry is
plotted in Fig.~5  for various form factors. The $Z$ boson mediated
diagrams give larger contributions than the photon mediated diagrams. 
The form factors $\hbox{ Im} f_6^Z$ gives the largest contribution. 
Therefore this measurement may be most effective in constraining 
$\hbox{Im } f_6^Z$ and less effective in constraining 
$\hbox{Im } f_4^\gamma$ and $\hbox{Im } f_6^\gamma$. 

One of the shortcomings of the asymmetry described above is that the angle 
$\Theta^{(>)}$ is defined with respect to 
one of the $W$ bosons.  Therefore one has to rely on good 
resolution on the $W$ boson momentum to define the angle.   That is not a 
necessity as far as detecting CP violation is concerned.  As analyzed 
earlier, the forward--backward asymmetry of the hardest jet itself, instead 
of the associated $W$ boson, is already a genuine CP--odd, 
CP$\hat{\hbox{T}}$--even 
observable.  Indeed this is a better observable to use in practice 
because it allows one to make less severe cuts in order to make sure 
the contributions due to the non-$W^+W^-$ source are gotten rid of.  
In fact, if the goal is simply to detect CP violation, it is not 
necessary to check whether the event is due to the $W$ boson or not.  
The forward--backward asymmetry of the hardest jet itself of any 
three-jets or four-jets events is already a genuine CP--odd, 
CP$\hat{\hbox{T}}$--even observable.   The $W^+W^-$ intermediate state is 
only one of the potential mechanisms giving rise to such an asymmetry.
It is only in the context of studying three gauge boson form factors that 
one wishes to eliminate the non--$W^+W^-$ background.  
To study the asymmetry in the context of $W^+W^-$ intermediate state, 
one has to go back to the general equation for cross section, 
$\sigma(\Theta, \psi, \phi, \bar\psi, \bar\phi)$, in 
Eq.~(\ref{eq:crosssection}).  For a given $\Theta$ the resulting jet can be 
either forward or backward.  Given the angles $\psi$ and $\phi$ of a parton 
the criterion for a jet to be forward is 
\begin{equation}
f(\Theta, \psi) = \gamma{\beta + \cos\psi \over \sin\psi} \cot\Theta > \cos\phi.
\end{equation}
One can use this criterion to integrate over the region of phase space for 
fixed $\Theta$ for which the hardest jet is forward and subtract those region 
in which the hardest jet is backward.  That is, 

\begin{eqnarray}
A_{FB}(\Theta) &=&
\int_{-1}^1 d\zeta\int_{-1}^1 d\bar\zeta 
\Bigl(\int_F - \int_B \Bigr) d\phi\int_0^{2\pi} d\bar\phi 
\; \sigma \; \vartheta(|\zeta|-|\bar\zeta|)                      \nonumber\\
&+& \int_{-1}^1 d\zeta\int_{-1}^1 d\bar\zeta 
\int_0^{2\pi} d\phi \Bigl(\int_{\bar F} - \int_{\bar B} \Bigr) d\bar\phi 
\; \sigma \;  \vartheta(|\bar\zeta|-|\zeta|) \;,
\end{eqnarray}
where F indicates integration region in which $\cos\phi < f(\Theta, \psi)$ 
when $\zeta=\cos\psi > 0$, or, $\cos(\pi + \phi) < f(\Theta, \pi - \psi)$ 
when $\zeta < 0$ while B represent the rest of the $\phi$ integration region.
For $\bar F$ and $\bar B$, the function $f$ should be changed to 
$f(\Theta, \bar\psi) = \gamma(- \beta + \cos\bar\psi/ \sin\bar\psi) 
\cot\Theta > \cos\bar\phi$.  
The first term in the above equation is the contribution 
when the $W^-$ provides the hardest jet, while the second term is when the 
hardest jet is originated from $W^+$ decay.
The total forward backward asymmetry will 
be an integration over $\Theta$.
In practice, it is much more practical to study the distribution
by Monte--Carlo simulation.
In Fig.~6a,b, we plot the difference of event distributions versus the
cosine of the polar angle of the hardest jet.  The size of error bars
comes from the numerical fluctuation in the simulation.
Comparing this with the corresponding curve in Fig.~5, one finds that the 
asymmetries, even though different in the two cases, are quite similar with 
the one defined with respect to the hardest jet directly larger by about 
a factor of two.  This indicates that there is indeed no unwanted 
cancellation when one uses the second, more directly measurable 
definition of asymmetry.

Unfortunately, defining the angles based on the hardest jet will not expose the
dispersive parts of the form factors.  This is reflected in
Eq.~(\ref{eq:asyHJ}) where only the absorptive parts are relevant.
One can try to use the CP$\hat{\hbox{T}}$-even asymmetry, $A_r$, defined 
earlier, however, it is an awkward observable to use.
 
As an alternative, one can avoid the reference to the polar angles in the $W$
decays altogether.  Then the system of $e^- e^+ \rightarrow W^-W^+ 
\rightarrow$ (four jets) 
looks like an unoriented production plane
and two decay half-planes.  Each half-plane extends in the direction of
the $W$ boson with which it is associated.  
One can define an angle 
$\varphi$ between a half-plane and the production plane by 
rotating a half-plane along its 
axis (defined by the direction of the momentum of its $W$) 
in a right-handed sense until it
coincides with the production plane.  
The angle of this rotation defines $\varphi$ taking value in $(0, \pi)$ 
in a unique way.  
There are two such azimuthal angles corresponding to the two $W$
bosons, and they must be distinguished.  By convention, one can define
the azimuthal angles so that $\varphi$ corresponds to the $W$ emitted in
the forward direction with respect to the incoming electron momentum.
The $\varphi'$ corresponds to the azimuthal angle in the jets of the 
recoiling $W$.
 
To conform to previous definition of the azimuthal angle, we shall also 
define $\varphi'$
by the right-handed rule along the direction of the forward $W$
just as in the definition of $\varphi$.
The $W^+W^-$ production angle $\Theta_{JP}$ in this case, by definition, 
only extends from 0 to
$\pi/2$, and both $\varphi$, $\varphi'\in (0,\pi)$. 
This coordinate system will be referred to as the Jet-Plane (JP) system.
The decay polar angles will be integrated out.  As before, a detailed
simulation should set the threshold variable $\epsilon$ to reflect the
experimental jet energy and direction resolution.  Here, it is assumed
that only two tiny holes in the phase space are cut out corresponding to
polar angles of 0 and $\pi$.
 
Now one can ask what the effect of a CP transformation is on a system
with this configuration.
One can convince oneself that under CP
\begin{equation}
\hbox{CP}:     \quad  (\Theta_{JP},\varphi,\varphi')\rightarrow
                      (\Theta_{JP},\varphi',\varphi).
\end{equation}
Under CP$\hat{\hbox{T}}$, 
\begin{equation}
\hbox{CP}\hat{\hbox{T}}
          :     \quad  (\Theta_{JP},\varphi,\varphi')\rightarrow
                      (\Theta_{JP},\pi-\varphi',\pi-\varphi).
\end{equation}

Define $n(\Theta_{JP},\varphi,\varphi')$ to be the differential rate
in JP variables.  At the quark level, 
the rate is $\sigma_q(\Theta,\phi,\bar\phi)$ for a
$W^-$ at a scattering angle $\Theta$
with the $W^-$ decay azimuthal angle $\phi$ and
the $W^+$ decay azimuthal angle $\bar\phi$.
A summation must be made over quark configurations
which lead to the same jet angles, $\Theta_{JP}$, $\varphi$, and
$\varphi'$.
\begin{eqnarray}
    n(\Theta_{JP},\varphi,\varphi') &=&
    \bigl[ \sigma_q(\Theta_{JP},\varphi,\varphi')
 +  \sigma_q(\Theta_{JP},\varphi+\pi,\varphi')\nonumber\\
&+& \sigma_q(\Theta_{JP},\varphi+\pi,\varphi'+\pi)
 +  \sigma_q(\Theta_{JP},\varphi,\varphi'+\pi) \bigr] \nonumber\\
&+& \bigl[ \Theta_{JP}\to\pi-\Theta_{JP}, \varphi \to -\varphi',
                               \varphi'\to -\varphi \bigr]
\;.
\end{eqnarray}
The last summing bracket interchanges the $W^+$ and $W^-$.
One way of constructing the integrated 
CP--odd asymmetry is to split the range of $\varphi$ 
into two quadrants and define 
the following asymmetry:
\begin{equation}
A_{JP}(\Theta_{JP})=
\frac{n(\Theta_{JP},\hbox{I,II})-n(\Theta_{JP},\hbox{II,I})}
     {n(\Theta_{JP},\hbox{all,all})},
\end{equation}
where I and II refer to the two quadrants of the azimuthal spaces
for both $\varphi$ and $\varphi'$.  Note that this asymmetry is 
CP$\hat{\hbox{T}}$-even and therefore probes only the dispersive parts 
of the form factors.  
Fig.~7 shows the numerical result for various form factors.
The form factors $\hbox{Re } f_6^\gamma$, $\hbox{Re } f_7^\gamma$ and 
$\hbox{Re } f_7^Z$ give rise to larger (positive) asymmetry than the 
other three form factors especially for $\cos\Theta_{JP} < 0.5$.
The numerator above is half of the following expression
$$
  n(\Theta_{JP},\hbox{I,all})  +n(\Theta_{JP},\hbox{all,II})
 -n(\Theta_{JP},\hbox{II,all}) -n(\Theta_{JP},\hbox{all,I}) \;, 
$$
which has the same form as the numerator in Eq.(\ref{eq:Appud}).
Therefore, we have the relation,
\begin{eqnarray}
&\ &
\frac{n(\Theta_{JP},\hbox{I,II})-n(\Theta_{JP},\hbox{II,I})}
     {n(\Theta_{JP},\hbox{all,all})}
=
{1\over\sigma(\Theta_{JP})+\sigma(\pi-\Theta_{JP})}\nonumber\\
\quad
\times
 \Biggl[&\ &
 {\sigma(\Theta_{JP}) \over\pi}
   \Bigl(\hbox{Im } \rho(\Theta_{JP})_{-,+}
    -\hbox{Im } \bar\rho(\Theta_{JP})_{+,-} \Bigr)              \\
\quad
&+&{\sigma(\pi-\Theta_{JP})\over\pi}
   \Bigl(\hbox{Im } \rho(\pi-\Theta_{JP})_{-,+}
    -\hbox{Im } \bar\rho(\pi-\Theta_{JP})_{+,-} \Bigr)
 \Biggr]
\;.                                               \nonumber
\end{eqnarray}
Here we have used the result in Eq.(\ref{eq:Apd}) to express the asymmetry 
in terms of the density-matrix elements. The differential cross section
$\sigma(\Theta)$ in the weighting factors is just the cross section for 
the $W$ pair production at $\Theta$.
One may feel that it is very difficult to reconstruct the 
on--shell $W$ bosons kinematically in order to partition the four jets into 
two pairs.  However, the accurate reconstruction of the $W$ bosons is 
in fact not really necessary.  This is because the hardness of the jets is 
purely due to the boost of the $W$ bosons as long as the quarks in the
final state are light.  Therefore, the $W$--partner of the hardest jet 
must be the softest one among the four.  
One can indeed apply this analysis to {\it any} 
$e^-e^+ \rightarrow$ four--jets 
events by pairing simply the hardest jet with the softest one in setting 
up one of the ``decay planes'' for the definition of the $\Theta_{JP}$.
Of course in that case, one cannot conclude that the CP--odd 
signal is purely due to the $W$ boson pair in the intermediate state.  
However, if the main goal is to look for any CP violating signal, this
may be an advantage instead of a set--back. 
 
\def\dump{
We now calculate the differential asymmetry as a function of the helicity
amplitudes.  To make our analysis more transparent we consider the production
alone, then add decay matrices later.
 
The differential cross section will be a sum over all the elements of the
density matrix ${\cal P}^{\lambda,\bar \lambda}_{\lambda,\bar \lambda'}
(\Theta)$
(see Ref.\cite{hagiwara}, Eq.4.11)
multiplied by some normalizing factors $N$ for the photon and $Z$
propagators, couplings, etc.
\begin{eqnarray}
{\cal P}^{\lambda,\bar \lambda}_{\lambda,\bar \lambda'}(\Theta) 
& = &\sum_{\sigma,\bar \sigma}
\tilde{\cal M}_{{\sigma,\bar \sigma},{\lambda,\bar \lambda}}
\tilde{\cal M}_{{\sigma,\bar \sigma},
{\lambda',\bar \lambda'} }^*;   \\
\frac{{\sigma}(\Theta)}{d\Theta} & = &N \sum_{{\lambda,\bar \lambda},
{\lambda',\bar \lambda'}}
{\cal P}^{\lambda,\bar \lambda}_{\lambda,\bar \lambda'}(\Theta)
\end{eqnarray}
 
If the $W$'s were stable and one could measure their helicity without their
decay one could consider cross sections for specific helicities:
\begin{equation}
{\sigma}(\Theta,{\lambda,\bar \lambda})/d\Theta 
= N {\cal P}^{\lambda,\bar \lambda}_{\lambda,\bar \lambda}
(\Theta).\:\:(No\;sum)
\end{equation}
 
The production matrices have useful symmetry properties.  Under
${\lambda,\bar \lambda} \rightarrow 
{\lambda',\bar \lambda'}$, $\cal P \rightarrow {\cal P}^*$.  Under $C$,
$\lambda \leftrightarrow \bar \lambda$ and
$\lambda' \leftrightarrow \bar \lambda'$.  Helicity is parity odd so under $P$,
all helicities are negated.   Therefore, under CP
\begin{equation}
{\cal P}^{\lambda,\bar \lambda}_{\lambda',\bar \lambda'} 
\begin{array}{r} CP \\ \rightarrow \end{array}
{\cal P}^{-\bar \lambda, - \lambda}_{-\bar \lambda', - \lambda'}
\end{equation}
 
From the CP properties of the helicity amplitudes as we have defined them, when
only dispersive parts of the form factors are present, the imaginary parts of
the production matrix are CP--odd and the real parts are CP--even.
 
Now we add the decay matrices for the semi-leptonic decay of the
$W$\cite{hagiwara}.  The decay matrices are
\begin{equation}
{\cal D}^{\lambda}_{\lambda'}=l_{\lambda}l_{\lambda'}^*
\end{equation}
and
\begin{equation}
{\bar {\cal D}}^{\bar \lambda}_{\bar \lambda'}={\bar l}_{\bar \lambda}
{\bar l}_{\bar \lambda'}^*
\end{equation}
where $l$ and $\bar l$ are, up to normalizing factors, the angular dependence of
the decay  amplitude for the $W^-$ and $W^+$ respectively.  With these
definitions the differential cross section for production followed by double
semileptonic decay is proportional to
\begin{equation}
\frac{{\sigma}(\Theta,{\lambda,\bar \lambda})}
     {d\Theta d\theta d\phi d\bar\psi d\bar
\phi}\propto  {\cal P}^{\lambda,\bar \lambda}_{\lambda',\bar \lambda'}
(\Theta)
{\cal D}^{\lambda}_{\lambda'}(\theta,\phi)
{\bar {\cal D}}^{\bar \lambda}_{\bar \lambda'}(\bar\psi, \bar \phi),
\end{equation}
(we sum over helicities unless otherwise specified).
If only one lepton is observed the differential cross section is given by
\begin{equation}
\frac{{\sigma}(\Theta,{\lambda,\bar \lambda})}
       {d\Theta d\theta d\phi}\propto
{\cal P}^{\lambda,\bar \lambda}_{\lambda',\bar \lambda}(\Theta)
{\cal D}^{\lambda}_{\lambda'}(\theta,\phi).
\end{equation}
For more detailed formulae, see Ref \cite{hagiwara}.
 
We integrate the decay matrices over regions of azimuthal space ($\phi, \bar
\phi$) to study their helicity structure.  First we divide the two azimuthal
domains in half.  This breaks the integrations down into events with leptons
(anti--leptons) above and below the $W^-W^+$ production plane.  The $\phi$
dependence of the decay matrix $\cal D$ is  $exp(i(\lambda - \lambda')\phi)$
and the $\bar \phi$ dependence of  $\bar {\cal  D}$ is $exp(i({\bar \lambda}' -
{\bar \lambda})\bar \phi)$.  Thus integrating over different azimuthal regions
selects different values of $\Delta \lambda \equiv (\lambda - \lambda')$ and
$\Delta {\bar \lambda}\equiv ({\bar \lambda}' - {\bar \lambda})$; when the
entire range is integrated,  the exponentials act like kronecker delta
functions in these helicity differences.
 
CP violation at the production vertex can only manifest itself in the energy
distribution of the leptons after this integration.
 
If we now halve our integration regions to $[0,\pi]$ and $[\pi,2\pi]$
for the azimuthal angles, we find that there
are contributions from $\Delta \lambda,\Delta {\bar \lambda}=0,\pm 1$:
\begin{eqnarray}
 \int_0^{\pi} d\phi {\cal D}^{\lambda}_{\lambda'}& = &
 (-1)^{\lambda-\lambda'} d_{-\lambda}(\theta) d_{-\lambda'}(\theta)
\left[
\pi \delta^{\lambda}_{\lambda'} \pm 2 i \delta^{\lambda}_{\lambda' \mp 1}
\right]
\\
&=&\left( \int_{\pi}^{2\pi} d\phi {\cal D}^{\lambda}_{\lambda'} \right)^*
\\
 & \equiv & X^{\lambda}_{\lambda'}(\theta)
\\
 \int_0^{\pi} d{\bar \phi} { \bar {\cal D}}^{\bar \lambda}_{\bar \lambda'}& = &
 (-1)^{\bar \lambda - \bar \lambda'}
\bar d_{-\bar \lambda}(\bar\psi) \bar d_{-\bar \lambda'}(\bar\psi)
\left[ \pi \delta^{\bar \lambda}_{\bar \lambda'}
\mp 2 i \delta^{\bar \lambda}_{\bar \lambda' \mp 1} \right]
\\
 & = & \left( \int_{\pi}^{2\pi} d\bar \phi
{\bar {\cal D}}^{\bar \lambda}_{\bar \lambda'} \right)^*
\\
 & \equiv & \bar X^{\bar \lambda}_{\bar \lambda'}(\bar\psi)
\end{eqnarray}
When $W$ decays into the two hemispheres of $\phi$ and $\bar \phi$ are
distinguished, CP and CP$\hat{\hbox{T}}$\ become distinguishable transformations
and the real parts of the form factors become accessible.  First we wish to
show consistency of this result with the statements about the charge-energy
asymmetry. A cross section
after integration over all four hemispheres will be proportional to
\begin{equation}
(X^{\lambda }_{\lambda'}(\Theta) +
X^{\lambda *}_{\lambda'}(\Theta))
({\bar X}^{\bar \lambda }_{\bar \lambda'}(\Theta) +
{\bar X}^{\bar \lambda *}_{\bar \lambda'}(\Theta))
\end{equation}
which is real and symmetric with respect to interchange of
primed and unprimed helicities.  Therefore,
when it is contracted with the production matrix only the real part will be
selected.  From Table 1 and our previous statements about the production
matrix, only CP violation due to absorptive parts of form factors is relevant.

Let's now consider an asymmetry in which we count all events with both lepton
and anti--lepton emitted above the production plane and subtract the number of
events where both lepton and anti--lepton are 
emitted below the production plane.
In contrast to the case above, when two hemispheres are integrated (UP and UP),
the cross section is proportional to
\begin{equation}
X^{\lambda }_{\lambda'}(\Theta)
{\bar X}^{\bar \lambda }_{\bar \lambda'}(\Theta)
\end{equation}
which is no longer symmetric in exchange of primed and unprimed helicities and
can be sensitive to the imaginary part of the production matrix (real part of
the CP violating form factors.  If we write
the product as a sum of real and imaginary parts we see that the real part is
CP--even and, ignoring absorptive parts of CP conserving form factors,
will not contribute to a CP--odd asymmetry of this type.
Thus this UP-UP asymmetry measures the real parts of CP non-conserving form
factors, though $f_7$ is not probed because it requires
$\Delta \lambda,\Delta {\bar \lambda}=\pm 2$.  Further dividing the azimuthal
range into quadrants will make $f_7$ observable with this type of asymmetry.
 
This detailed analysis illustrates the way in which the rules relating CP
and CP$\hat{\hbox{T}}$\ are enforced.
}

\section{CP Asymmetry in the purely leptonic mode}

In the process $e^-e^+\to W^-W^+\to (\ell^-\bar\nu)(\ell'^+\nu)$, the two
missing neutrinos are not detectable. It is known that kinematics can
only be re--constructed with a two--fold ambiguity. To be 
self--contained, we shall explain this re--construction geometrically.
The vectors $\vec{p}_{W^-}$ and $\vec{p}_{W^+}$ lie on the surface of a
sphere $S$  of radius ${1\over2}(s-4M_W^2)^{1\over2}$. The direction of
$\vec{p}_{\ell^-}$ fixes a point $P$ on $S$. Since the magnitude the
missing $\vec{p}_{\bar\nu}$ is just 
$E_{\bar\nu} = \case1/2\sqrt{s} - E_{\ell^-}$, 
the angle $\theta$ between $\vec{p}_{\ell^-}$ and $\vec{p}_{W^-}$ 
is fixed to be 
$\cos\theta = (|\vec{p}_{\ell^-}|^2 + |\vec{p}_{W^-}|^2 - E_{\bar\nu}^2)/
|\vec{p}_{\ell^-}||\vec{p}_{W^-}|$.   
The closed triangle described by
the vectors $\vec{p}_{\ell^-}$, $\vec{p}_{W^-}$, and $\vec{p}_{\bar\nu}$
has a fixed shape, but it is free to swing about the axis
$\vec{p}_{\ell^-}$. 
Therefore, the locus of $ \vec{p}_{W_-}$ is a circle $C$ on the sphere
$S$ about point $P$. similarly, the locus of $-\vec{p}_{W_+}$ is another
circle $C'$ on the sphere $S$ about point $P'$, which corresponds to the
direction of $-\vec{p}_{\ell'^+}$. 
Since $W^+$ and $W^-$ are recoiling
one another, the solution of the physical directions of $W^-$ and $W^+$ 
are locations of the
intersections of these two circles $C,C'$. Generally, these two circles
have two common intersecting points, which give rise to the two folded
uncertainty that we cannot resolve without knowing the direction of each
neutrino momentum. 

If we wish to measure the up--down asymmetries in 
Eqs.(\ref{eq:Aud}--\ref{eq:Apd}), we need to know whether 
$\ell^-$ or $\ell'^+$ in each event is above or below the production plane.  
The two--folded ambiguity in fixing the $W^+W^-$ production plane
will not always prevent us from making such determination --- there 
is a region of phase space in which both solutions for the
production plane agree that a charged lepton was ``above'', or ``below'',
the production plane. However, some of the asymmetry will be averaged 
away by this ambiguity and relying on this mode of partial
reconstruction will reduce statistics. 

An alternative is to use CP violating observables which depend only on 
the observed momenta $\vec p_{e^-}$, $\vec p_{e^+}$, $\vec p_{\ell^-}$,
and $\vec p_{\ell'^+}$. For example, we can measure the overall asymmetry 
as follows,
\begin{equation}
\quad
{
 N(\vec p_{e^-} \cdot \vec p_{\ell'^+} \times \vec p_{\ell^-} >0)
-N(\vec p_{e^-} \cdot \vec p_{\ell'^+} \times \vec p_{\ell^-} <0)
\over
 N(\vec p_{e^-} \cdot \vec p_{\ell'^+} \times \vec p_{\ell^-} >0)
+N(\vec p_{e^-} \cdot \vec p_{\ell'^+} \times \vec p_{\ell^-} <0)
}
\;.
\end{equation}
This quantity is C-even, P-odd and $\hat{\hbox{T}}$-odd therefore only the 
real parts of the CP--odd form factors contribute.  
In Fig.~8a,b, we use Monte--Carlo simulation to illustrate this asymmetry 
in the event distribution versus the variable 
$\chi=(8/sM_W)\vec p_{e^-} \cdot \vec p_{\ell'^+} \times \vec p_{\ell^-}$.  
Only $\hbox{Re } f_{4,6}^{\gamma,Z}$ are used as illustrations.  
They show that $\hbox{Re } f_6^{\gamma,Z}$ generally give rise to larger 
contributions than $\hbox{Re } f_4^{\gamma,Z}$.

One can also understand this asymmetry geometrically by 
constructing two oriented planes. The first plane
is specified by $\vec p_{\ell^\pm}$.  
The other plane is constructed by the beam 
direction and the total missing momentum.
The relevant observable is the angle $\Phi$ between these two 
planes. CPV will appear in a way that the event distribution is not
symmetric under the transformation $\Phi \leftrightarrow -\Phi$.
Under CP$\hat{\hbox{T}}$, $\Phi$ is invariant.
In Fig.~9a,b, we use Monte--Carlo simulation to illustrate this asymmetry 
in the event distribution versus the variable $S = \sin\Phi$
using $\hbox{Re } f_4$ and $\hbox{Re } f_6$ as examples.
Just as in Fig.~8a,b, $\hbox{Re } f_6^{\gamma,Z}$ generally give rise to 
larger contributions than $\hbox{Re } f_4^{\gamma,Z}$.

Note that this CP violating observable can be applied to any process with 
$e^- e^+ \rightarrow \ell^- \ell'^+ X$ where $X$ is some neutral object.
The $W^+ W^-$ intermediate state is only one of the mechanisms that 
can give rise to this asymmetry in general.  
For example, another process that can contribute to this asymmetry is 
$e^- e^+ \rightarrow Z + Z^* \rightarrow \ell^- \ell^+ \nu \bar\nu$.  
More exotically, the signal may originate from wino pair production in 
supersymmetric models.  
If one wishes to investigate only the $W^+ W^- \gamma(Z)$ form factors one 
can do further kinematic cuts based on the determination of $W$ momenta 
up to a two-fold ambiguity.  Alternatively, one can insist that $\ell'$ 
and $\ell$ are of different flavor.  In that case the non--$W$--pair 
background can be greatly reduced.

\section{MODELS OF CP VIOLATION}
\label{sec:models}
 
In specific gauge models of CP violation, the natural values for the 
CP--odd form factors considered here are expected to be $10^{-2}$ or 
smaller due to the one loop suppression factor.  That makes observation of 
the CP--odd effects discussed here a very difficult task.  
In this section we briefly discuss several possibilities of having 
models which can lead to asymmetries which may be observable.  
As noted earlier, the standard, KM model is not expected to give 
significant CP violation signals at high energy colliders.  
One reason is that the CP violation will always be
proportional to the sines of the mixing angles, an a priori suppression
of about $10^{-3}$ over CP conserving processes, even at high energy. In
other models such as multi-Higgs doublet models, where CP violation
arises in the Higgs sector, CP violation is suppressed at low energy
because the mass of the Higgs bosons is so high.  At higher energies,
such suppression effect disappears.  
Similarly, one can also imagine SUSY models in which 
CP symmetry is broken in the couplings of heavy superpartners to gauge
bosons. 
 
Since $f_6$ and $f_7$ contain the parity violating $\varepsilon$ symbol, 
a general argument\cite{r:ckl} shows that it takes at least one fermion
loop to generate them even in non-standard models. Among the CP odd form
factors, $f_7$ is the one that is least likely to be generated at the
one loop level in non-standard models. To generate $f_7$ form factor, in
addition to the fermion loop, there have to be enough factors of momentum from
the fermion propagators to provide three powers of external momenta. To
achieve this, one has to avoid the mass insertion in the three fermion
propagators. It is easy to show that if the neutral gauge boson vertex
is flavor neutral, for the one loop graph to be CP violating, it
requires coexistence of the left the right handed couplings in the $W$
coupling vertex. To achieve CP violation, it is not possible
avoid mass a insertion in fermion propagators. Therefore, there are
insufficient momenta in one loop to generate the three powers required.  
In contrast, $f_4$ and $f_6$ typically can be generated at one loop in 
most non-standard models.  The parity--even 
$f_4$ can even be induced in purely bosonic loop graphs. 

He, Ma, and McKellar\cite{r:He} have calculated the one-loop
contributions to the CP--odd form factors in the two Higgs doublet and
left-right symmetric models. 
 
In the two Higgs doublet model, there are three neutral and one charged
Higgs boson which can all contribute from within the loop shown in
Fig.~10. He {\it et al.} find that only $f_4^Z$ is generated and that it
can be as large as $10^{-3}$.  Note that the presence of an absorptive
part in the form factor depends on the mass of the Higgs bosons; Higgs
bosons lighter than $\case1/2\sqrt{s}$ can go on shell in the loop. 
However, one also have to take into account the 
constraints on the masses in the Higgs sector
from flavor changing neutral current (FCNC) bounds.  To the degree that
the natural flavor conserving discrete symmetry is broken, CP
violation may appear in the the Higgs sector.  Therefore the constraint on 
FCNC also tends to limit CP violation as well.  
 
The Left-Right symmetric model is an example of a class of models in
which the CP violating form factors are generated at one loop by fermion
loops such as those in Fig.~11.  These models have have left- and
right-handed couplings to $W$ bosons each with a different phase.  To
generate $f_4$ the axial coupling is required, hence there is no
$f_4^{\gamma}$ induced.  To generate $f_6^{\gamma,Z}$ the vector coupling is
required. In a Left-Right symmetric model in which CP violation arises
in the mixing of the $SU(2)_L$ and $SU(2)_R$ gauge bosons, Ref
\cite{r:He} find that at $\sqrt{s}=200$~GeV, $f_4^Z \sim 10^{-4}$,
$f_6^Z \sim 10^{-4}$, and $f_6^{\gamma} \sim 10^{-3}$.  The mixing angle
is constrained by low energy physics\cite{pdb} to be {\it less than}
0.0028 and this puts $f_4^Z$ and $f_6^Z$ beyond LEP--II's limit of
detectability, and $f_6^{\gamma}$ is on the very edge. 
 
There are other models in this class which may be 
more promising because they are less constrained experimentally.  
For example,\cite{r:Fab} one can imagine that a subset of the 
supersymmetric spectrum consisting of winos and photinos with similar
couplings to those of the quarks in the Left-Right model generates a
larger $f_4^Z$, $f_6^Z$ and $f_6^{\gamma}$ (Fig.~11 replacing quarks with
$W$ and $\gamma$ superpartners). This is because there is no
constraint on the CP violating source in such models.  The masses of the
winos or photinos can be accommodated by current data in the range less
than $\case1/2\sqrt{s}$ and larger than the current bounds, thus
providing an absorptive part to the above form factors. 
 
\section{CONCLUSION}
\label{sec:conclusion}

At LEP II, the $W^+ W^-$ production cross section reach the maximum of
$\sigma \simeq 20 pb$ for $\sqrt{s} = 200$ GeV which can provide about
$10^4$ $W^+ W^-$ pairs per year for the design luminosity of $5 \cdot
10^{31} cm^{-2} sec^{-1}$.  The branching ratio of the $W$ decay into
each lepton channel is about ${1 \over 9}$ while each of the light quark
channel is about $\case1/3$.  This gives us a lot of events for both the
leptonic and the hadronic channels.  As we have shown in the paper, it
is possible to test CP symmetry in purely leptonic, purely hadronic or
and mixed channels of the two $W$ boson decays.  While the event
statistics probably will not be large enough to test some of the popular
alternative gauge models of CP violation, it is nevertheless sufficient
to provide nontrivial constraints on the CP--odd form factors in the
three gauge boson couplings. 

Even though in this paper we have concentrated on decoding the CP--odd 
form factors of the three gauge boson couplings in the process 
$e^- e^+ \rightarrow W^- W^+$, 
many of the techniques we used in the analysis 
can also be applied to other high energy collision processes which 
may be relevant to some special models of CP violation.  This is especially 
true for the observables we analyzed in Section V and VI.  It is interesting 
to measure the CP--odd signal in events with four jets or with two charged 
leptons and missing momentum in $e^-e^+$ colliders, independent of 
whatever the intermediate states that may be used to interpret such signals. 
In case of null signals, they can be translated into constraints about 
various form factors in different models.

\newpage
\centerline{Figure Caption}
\begin{itemize}
\item[Fig. 1]  The coordinate system for the process $e^-e^+\rightarrow
W^-W^+ \rightarrow 
({\mbox{f}}_1\bar{\mbox{f}}_2)({\mbox{f}}_3\bar{\mbox{f}}_4)$
Fig.1a is the $W$ pair production plane.  Fig.1b is the $W^-$ decay 
kinematics.  Note that the angles
$\psi$ and $\phi$ are defined {\em in the rest frame of the decaying $W$},
while the angle $\Theta$ is defined in the center of mass system.
The $x$, $y$, and $z$ axes which define the angles
$\bar\psi$ and $\bar\phi$ for the $W^+$ decay
also share the same directions as those depicted in the figure; 
again, the angles are defined in the rest frame of the $W^+$.
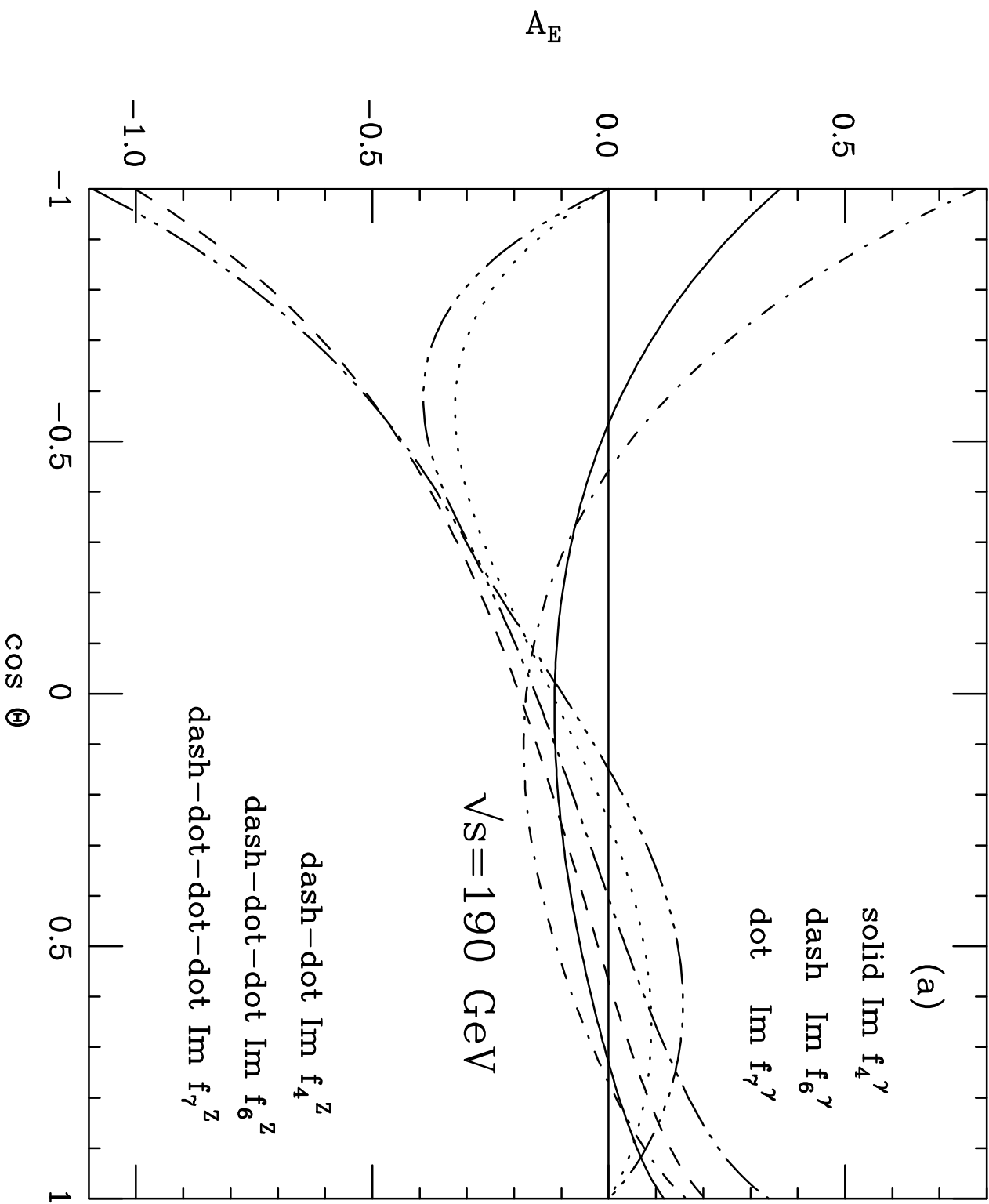
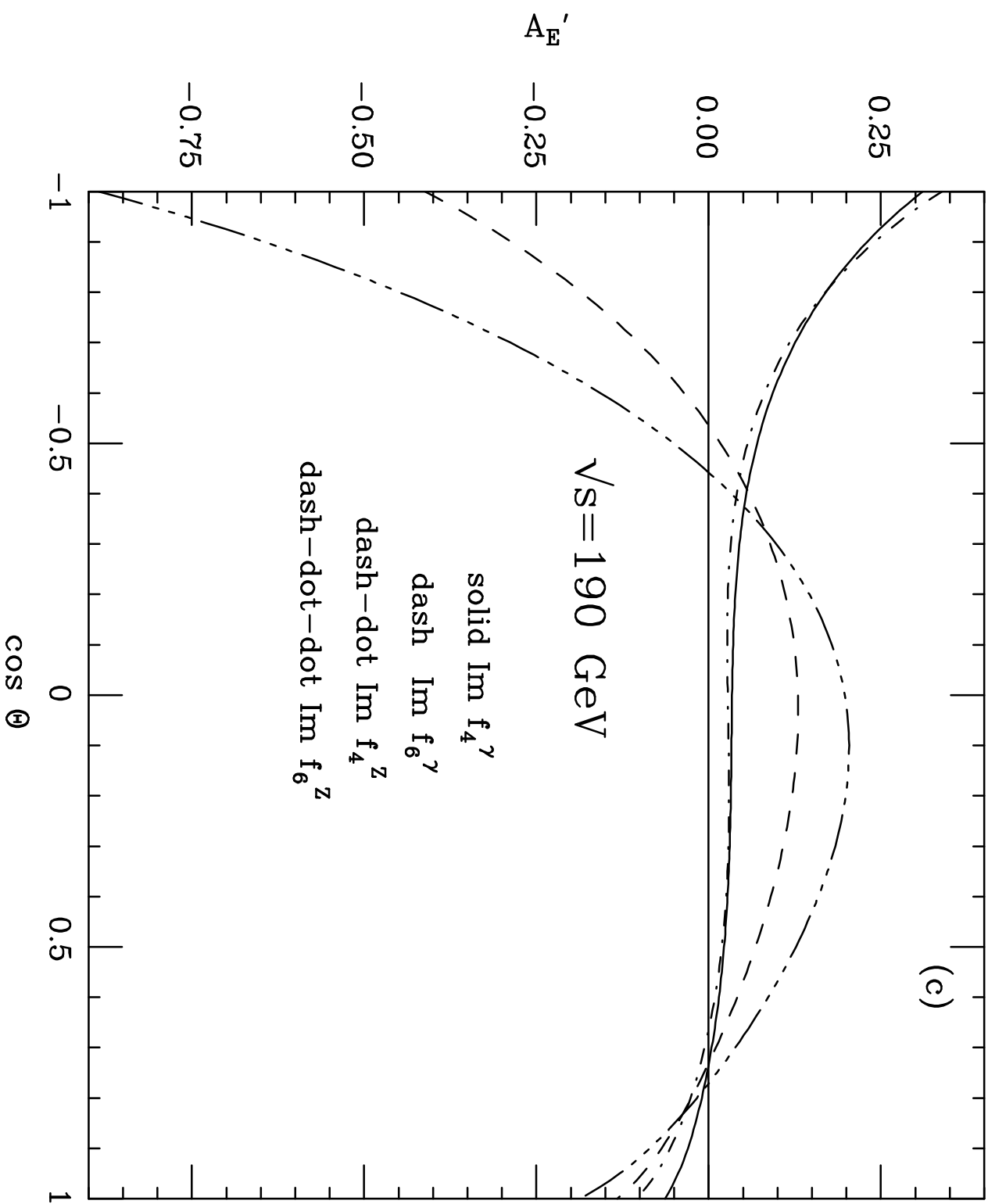
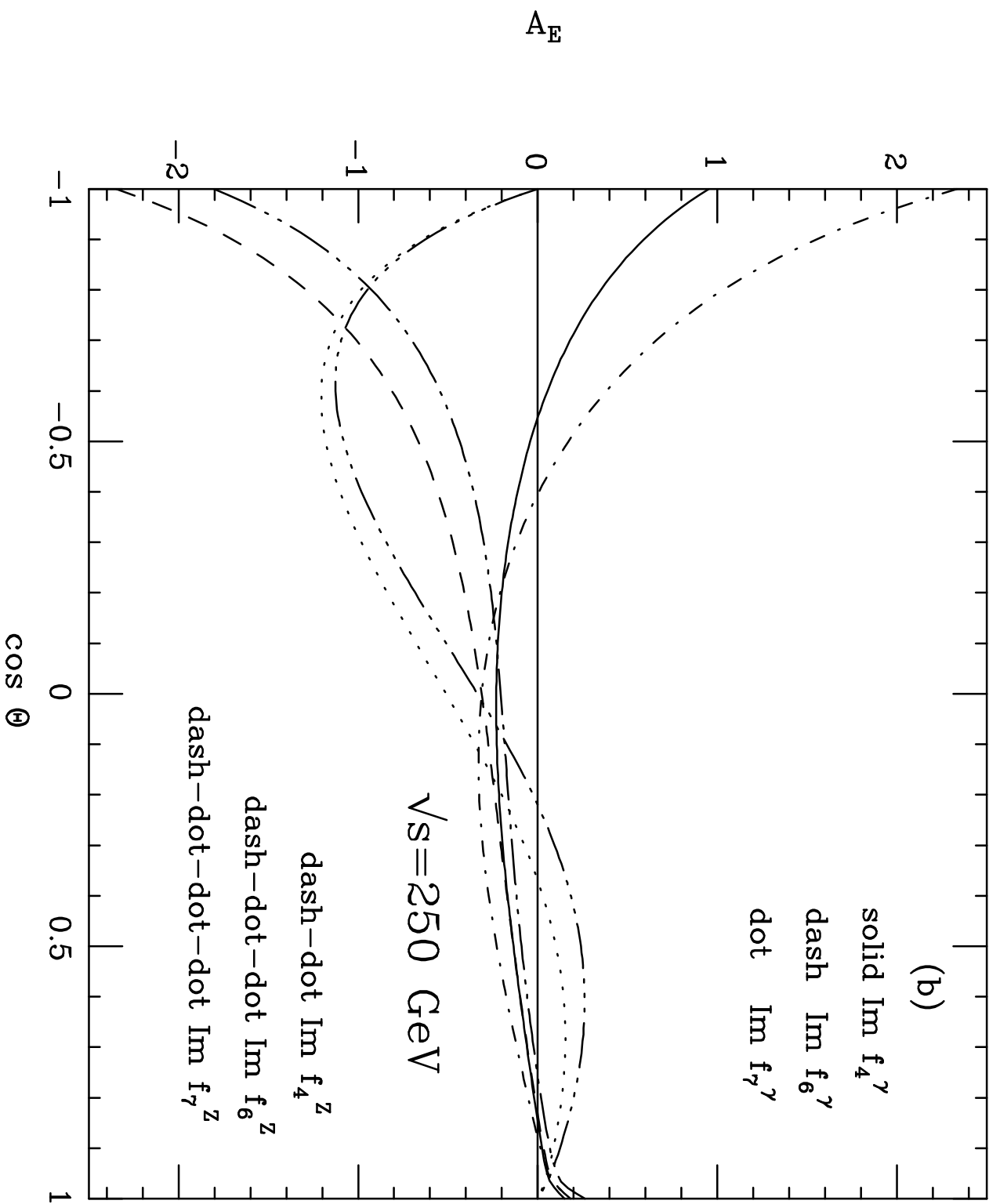
\item[Fig. 2] 
(a) $A_E$, per unit of Im $f$, at $\sqrt{s}=190$ GeV, 
(b) $A_E$, per unit of Im $f$, at $\sqrt{s}=250$ GeV, 
(c) $A'_E$ per unit of Im $f$, at $\sqrt{s}=190$ GeV.
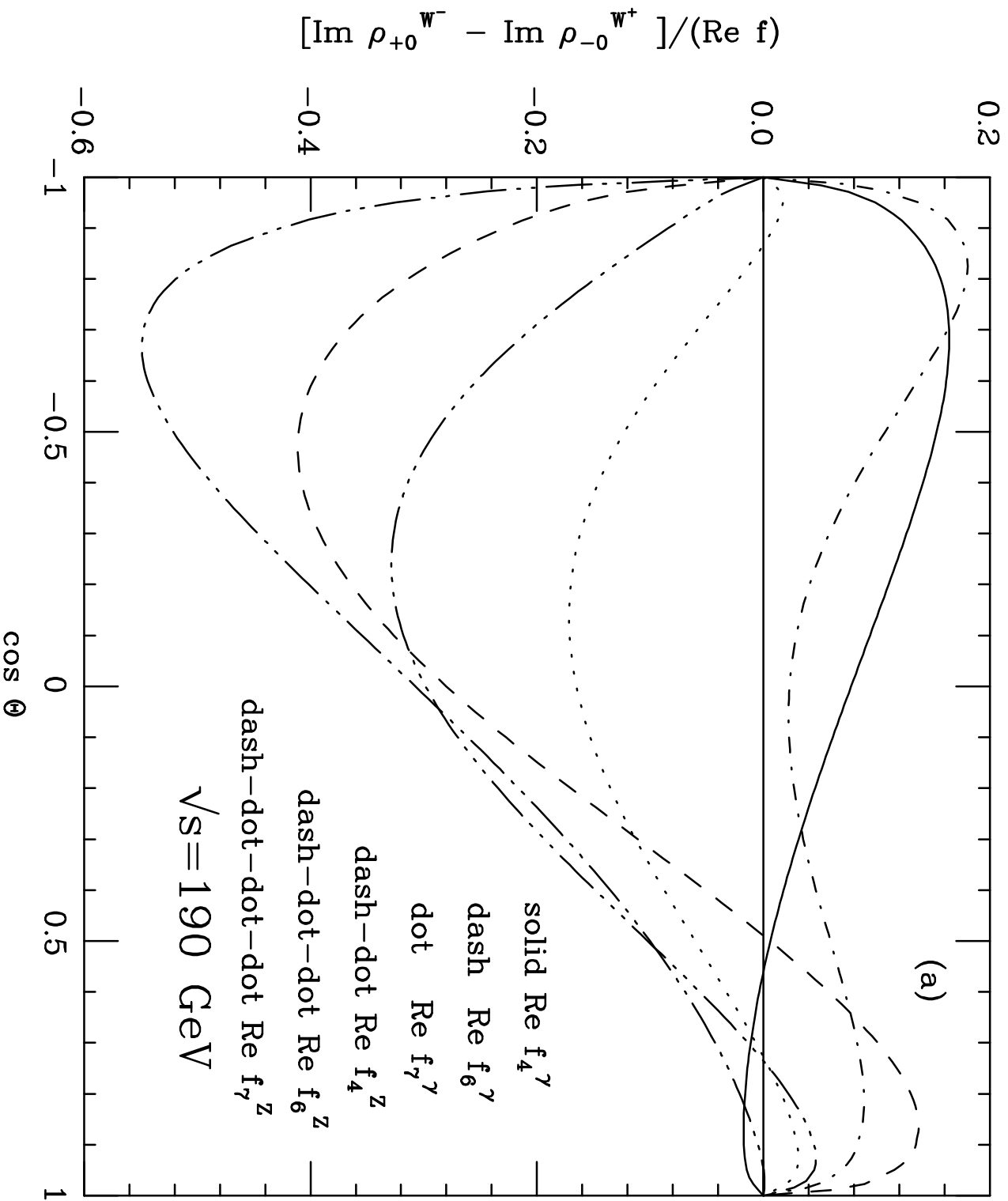
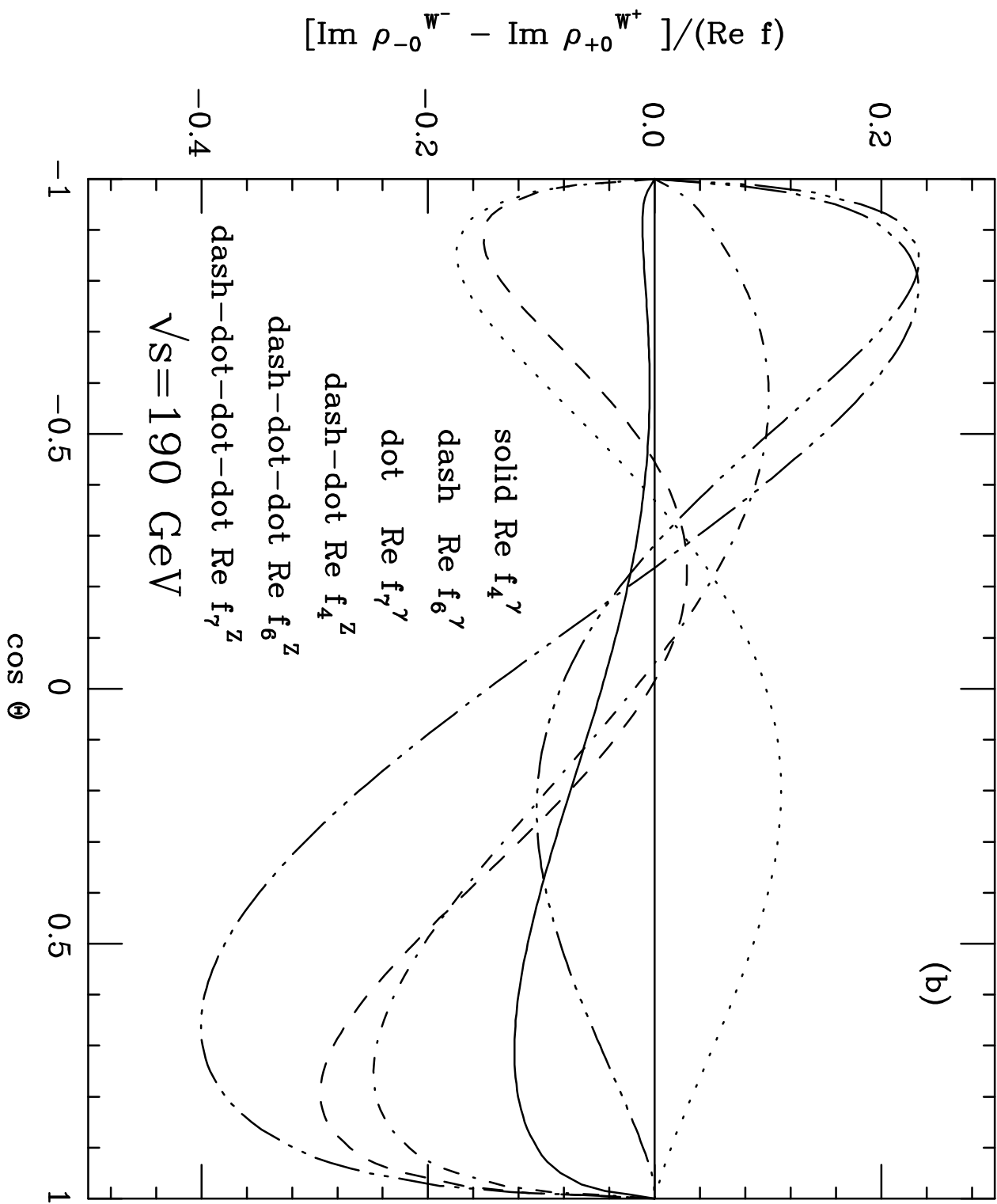
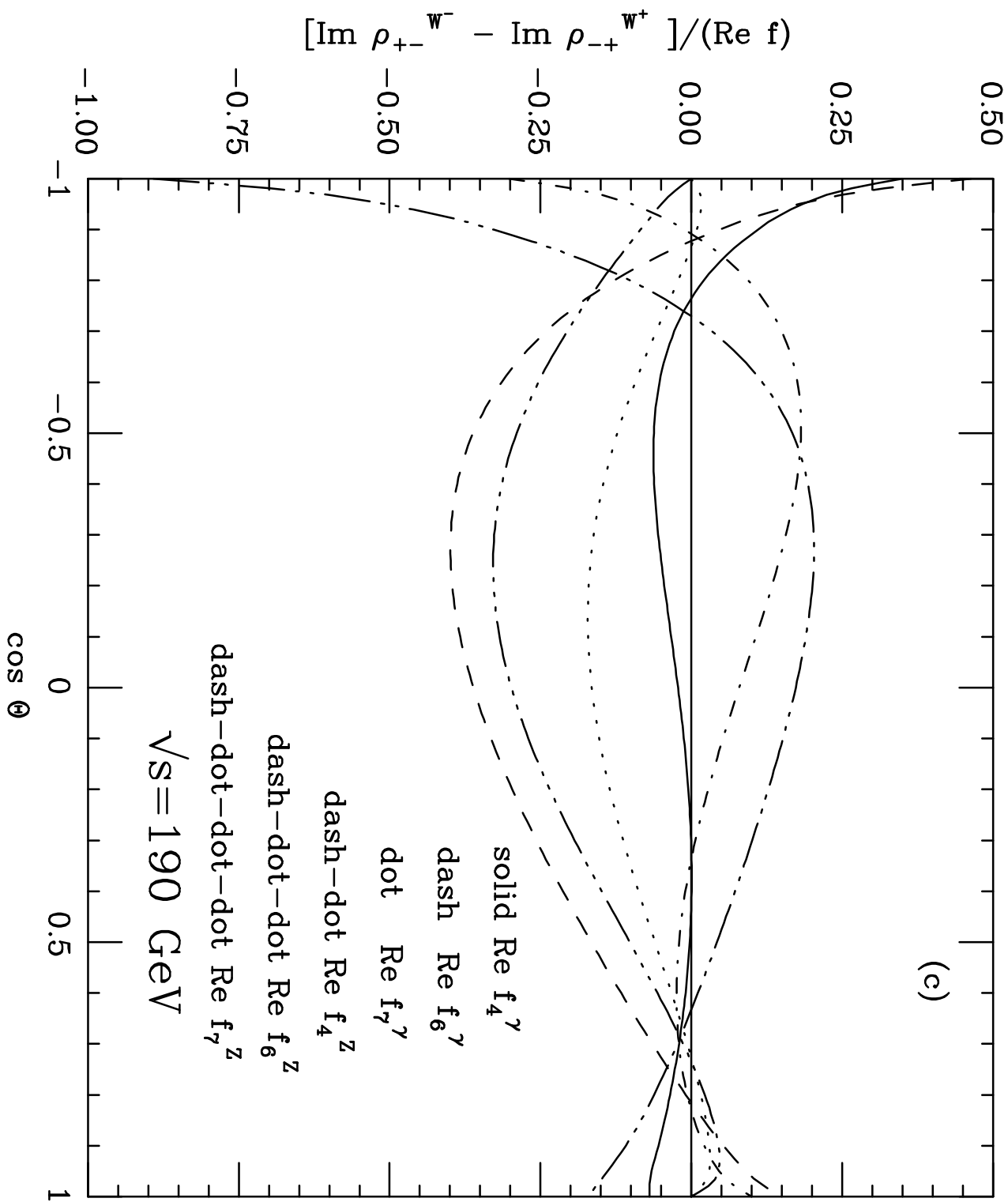
\item[Fig. 3] (a) Im $\rho_{+,0}-$Im $\bar\rho_{-,0}$,
(b) Im $\rho_{-,0}-$Im $\bar\rho_{+,0}$, and
(c) Im $\rho_{+,-}-$Im $\bar\rho_{-,+}$,  per unit of Re $f$,
versus $\Theta$ at $\sqrt{s}=190$ GeV.
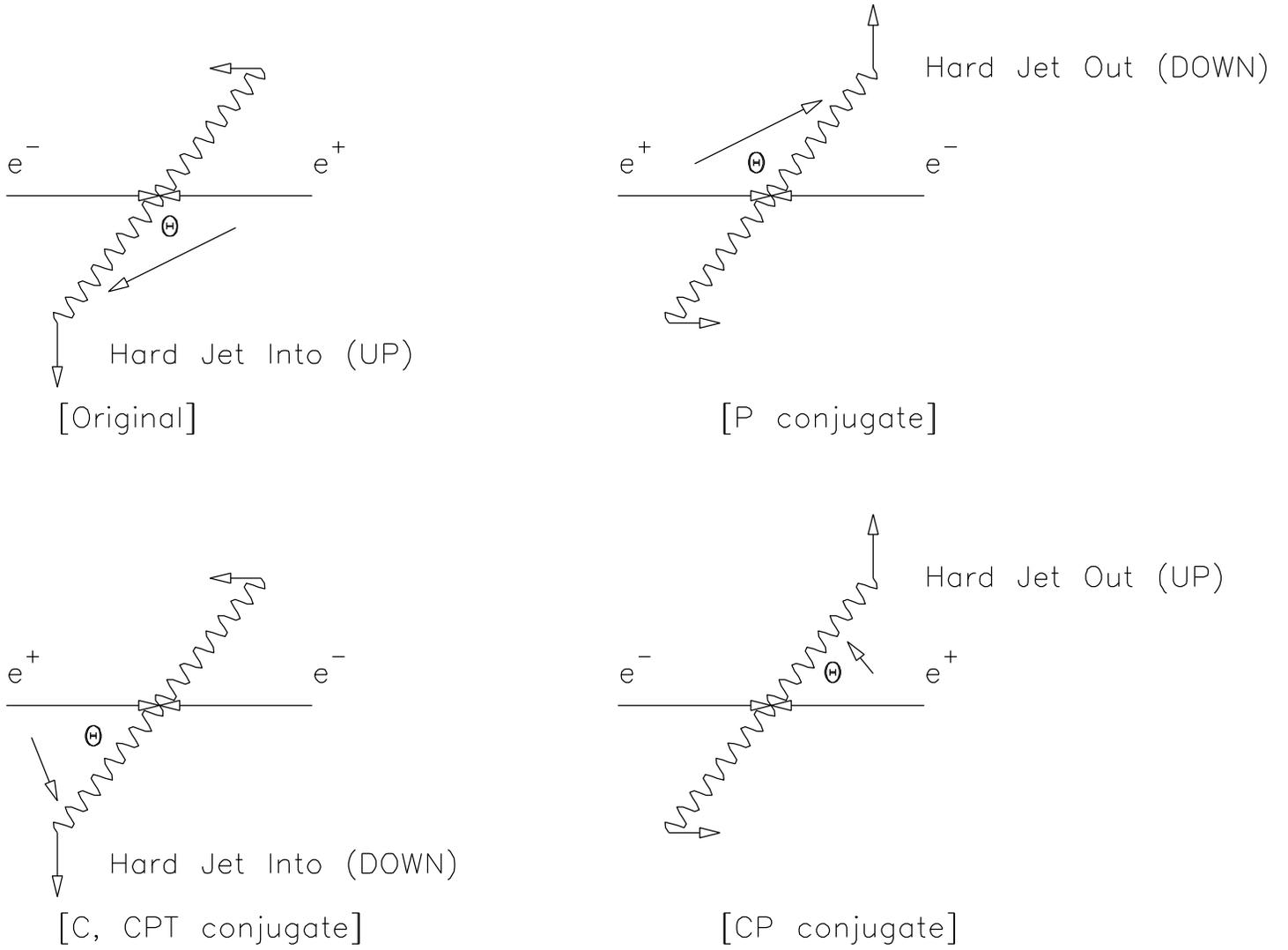
\item[Fig. 4] Discrete transformations of a typical
$e^-e^+\rightarrow W^-W^+ \rightarrow$  4--jet event.
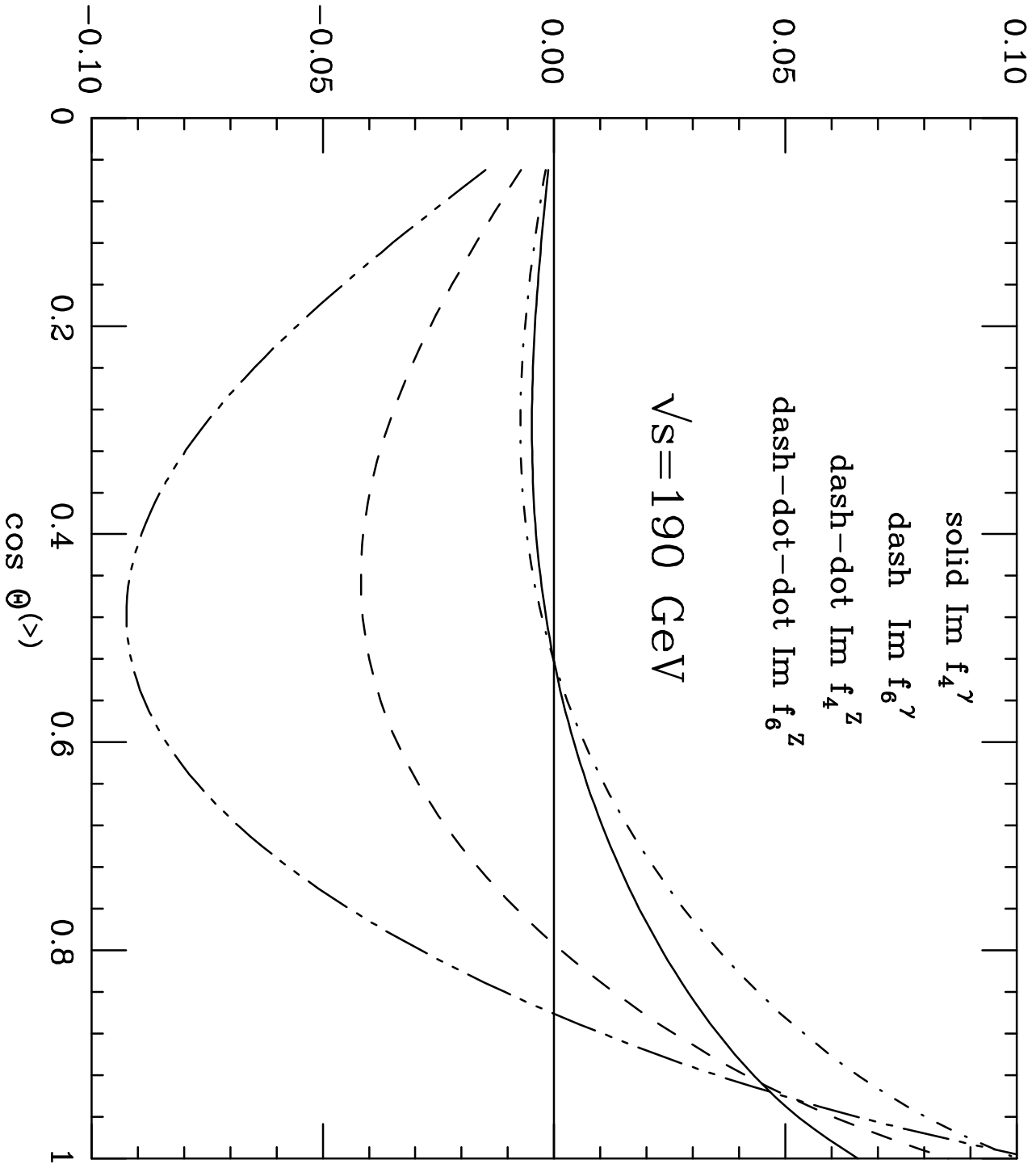
\item[Fig. 5] The forward--backward asymmetry of $W$ boson that is 
associated with the most energetic jet in
$e^-e^+\rightarrow W^-W^+ \rightarrow 4$--jets.  
Each curve corresponds to an
individual CPV form factor having an absorptive part 
while the others are set to zero.
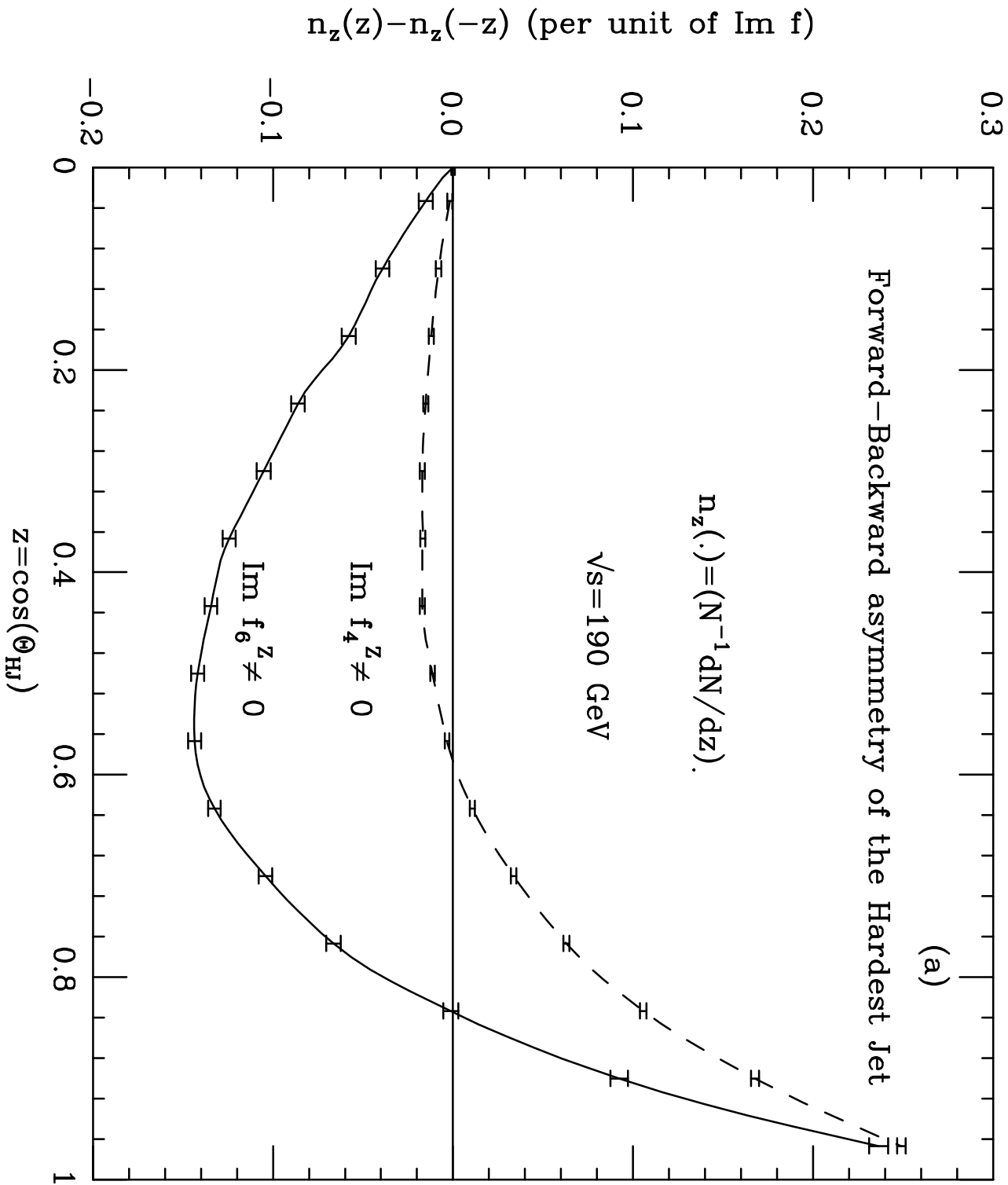
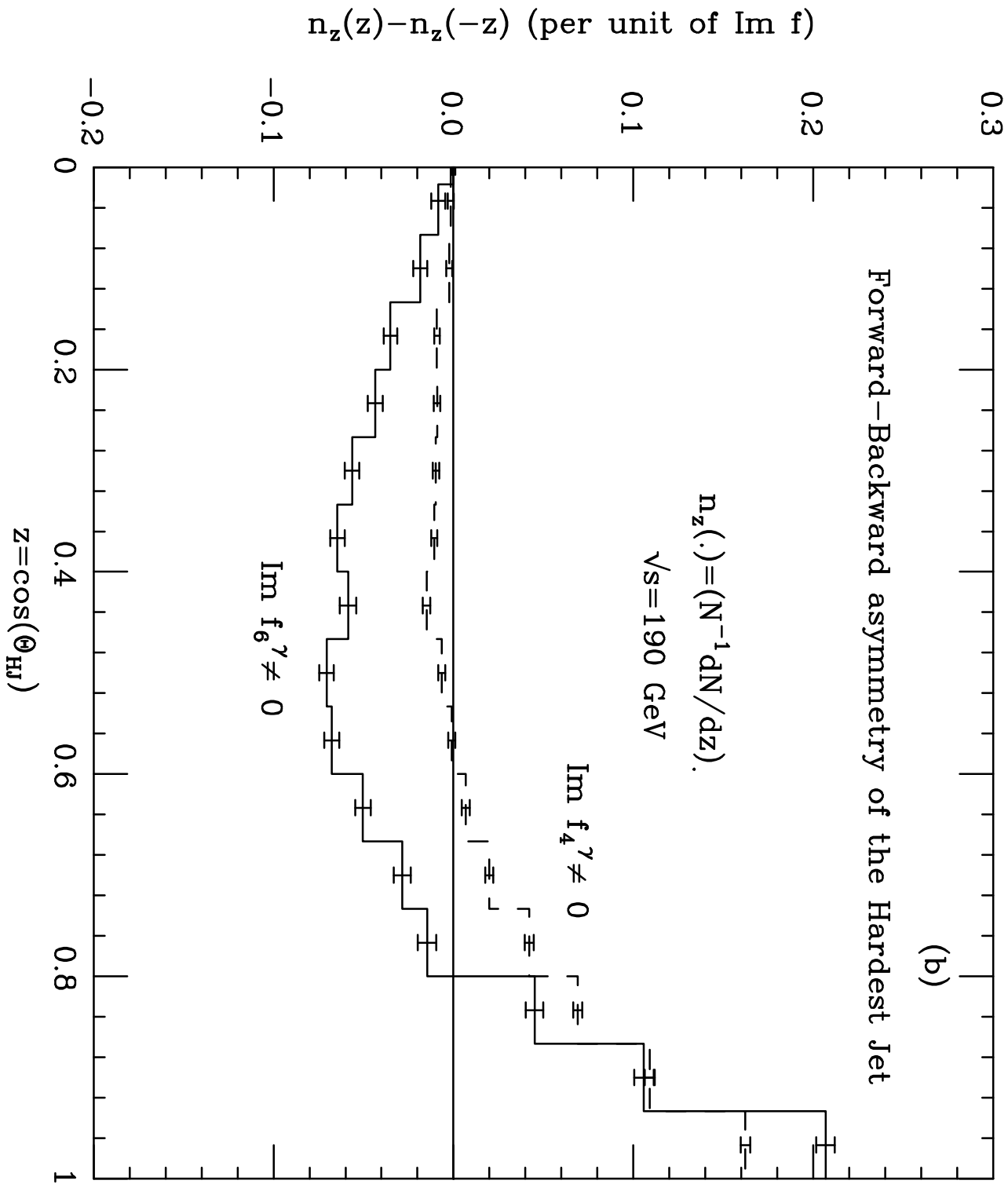
\item[Fig. 6]
(a) Result from Monte--Carlo simulation for the forward--backward
asymmetry of the most energetic jet itself in $e^-e^+\rightarrow W^-W^+
\rightarrow 4$--jets versus the cosine of the polar angle $\Theta_{HJ}$
of the jet.
Asymmetry here is measured per unit of $\hbox{ Im} f_4^Z$ 
or $\hbox{ Im} f_6^Z$ with other CPV form factors tunred off.
(b) Same as (a) for the form factors $\hbox{ Im} f_4^{\gamma}$ 
and $\hbox{ Im} f_6^{\gamma}$.  For readability, smooth Monte--Carlo results are
joined with a smooth curve.     
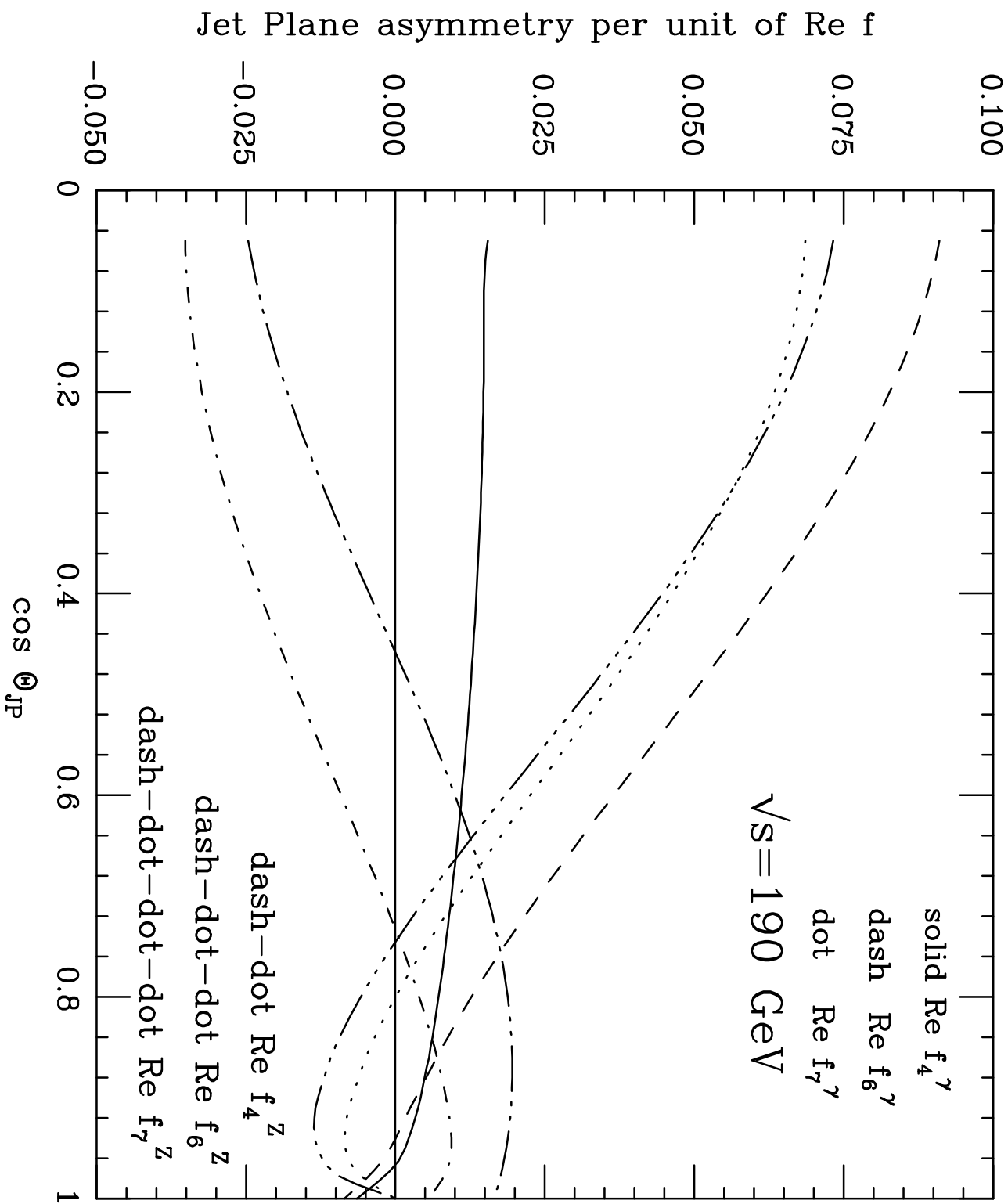
\item[Fig. 7]
The jet plane asymmetry for $e^-e^+\rightarrow W^-W^+ \rightarrow
4$-jets as a function of $\cos\Theta_{JP}$.  Each curve corresponds to
an individual form factor having an dispersive part while the others are
held at zero. 
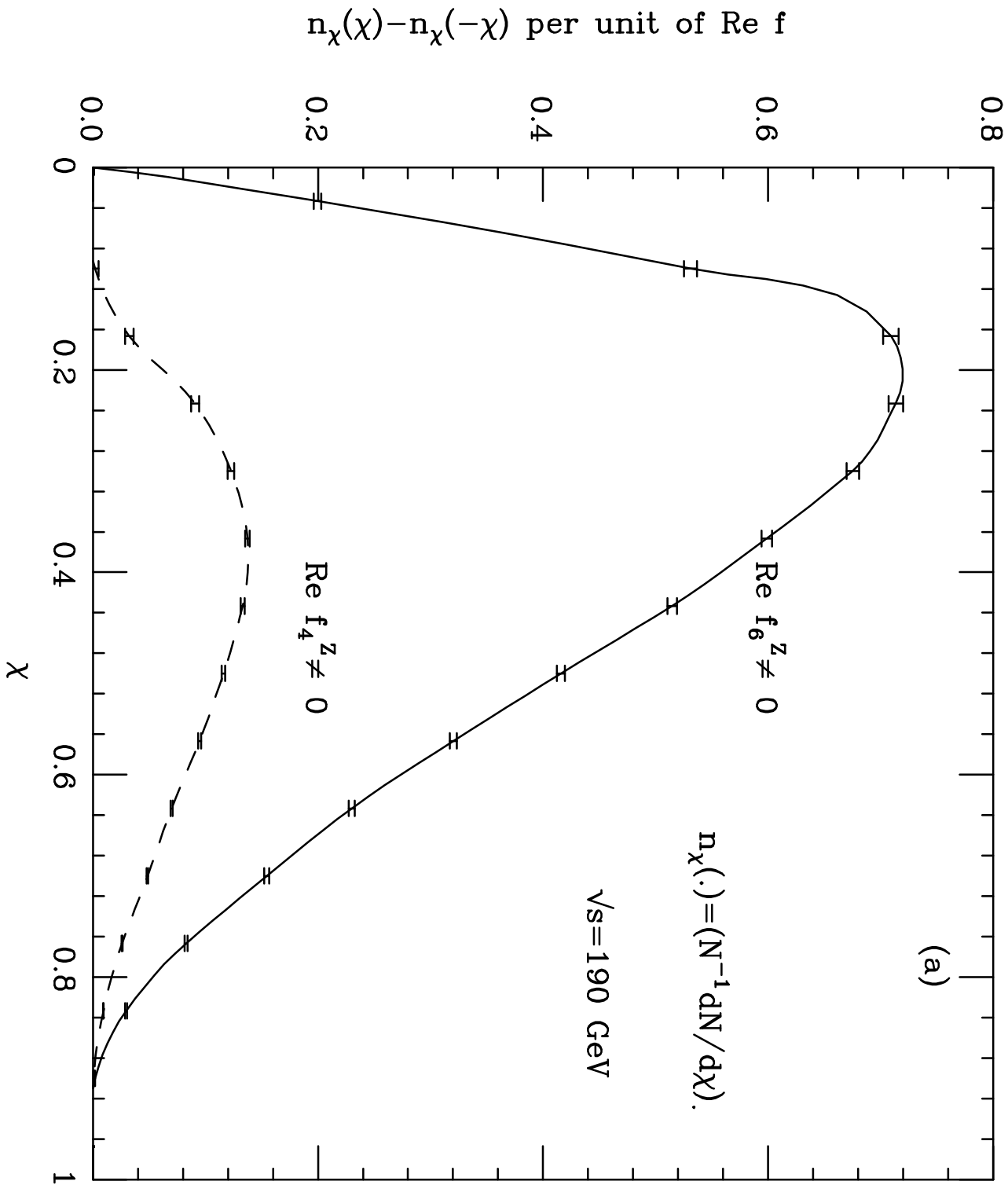
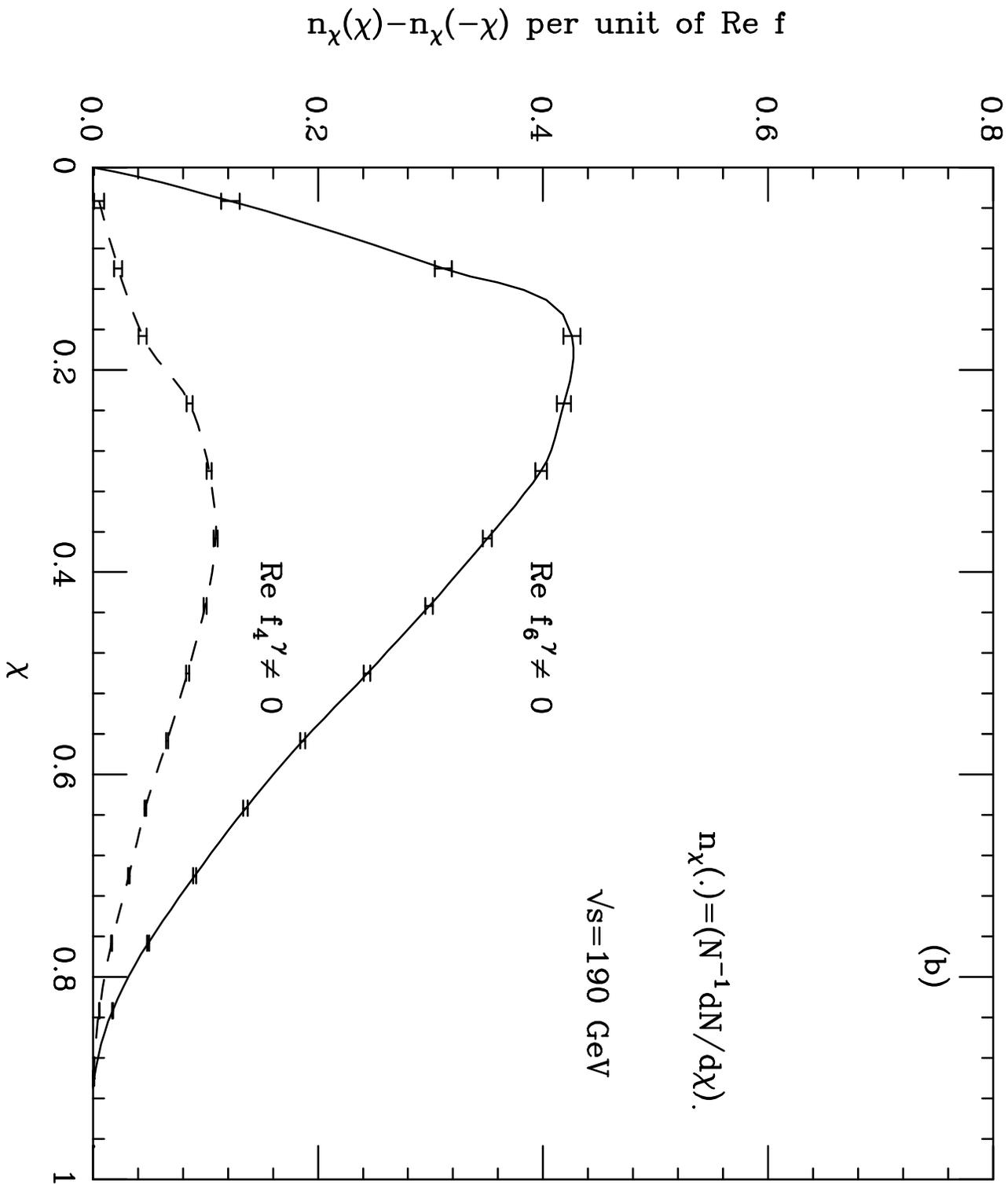
\item[Fig. 8]
(a) Result from Monte--Carlo simulation for the purely leptonic mode. The
distribution difference $n_\chi(\chi)-n_\chi(-\chi)$ is 
plotted at $\sqrt{s}=190$ 
GeV per unit of $\hbox{Re } f_4^Z$ or $\hbox{Re } f_6^Z$. Here
$\chi=(8/sM_W)\vec{p_{e^-}} \cdot \vec p_{\ell^+} 
\times \vec p_{\ell^-} $, and
$n_\chi(\chi)=N^{-1}dN/d\chi$.
(b) Same as (a) for the form factors 
$\hbox{Re } f_4^{\gamma}$ and $\hbox{Re } f_6^{\gamma}$. 
For readability, smooth Monte--Carlo results are
joined with a smooth curve.     
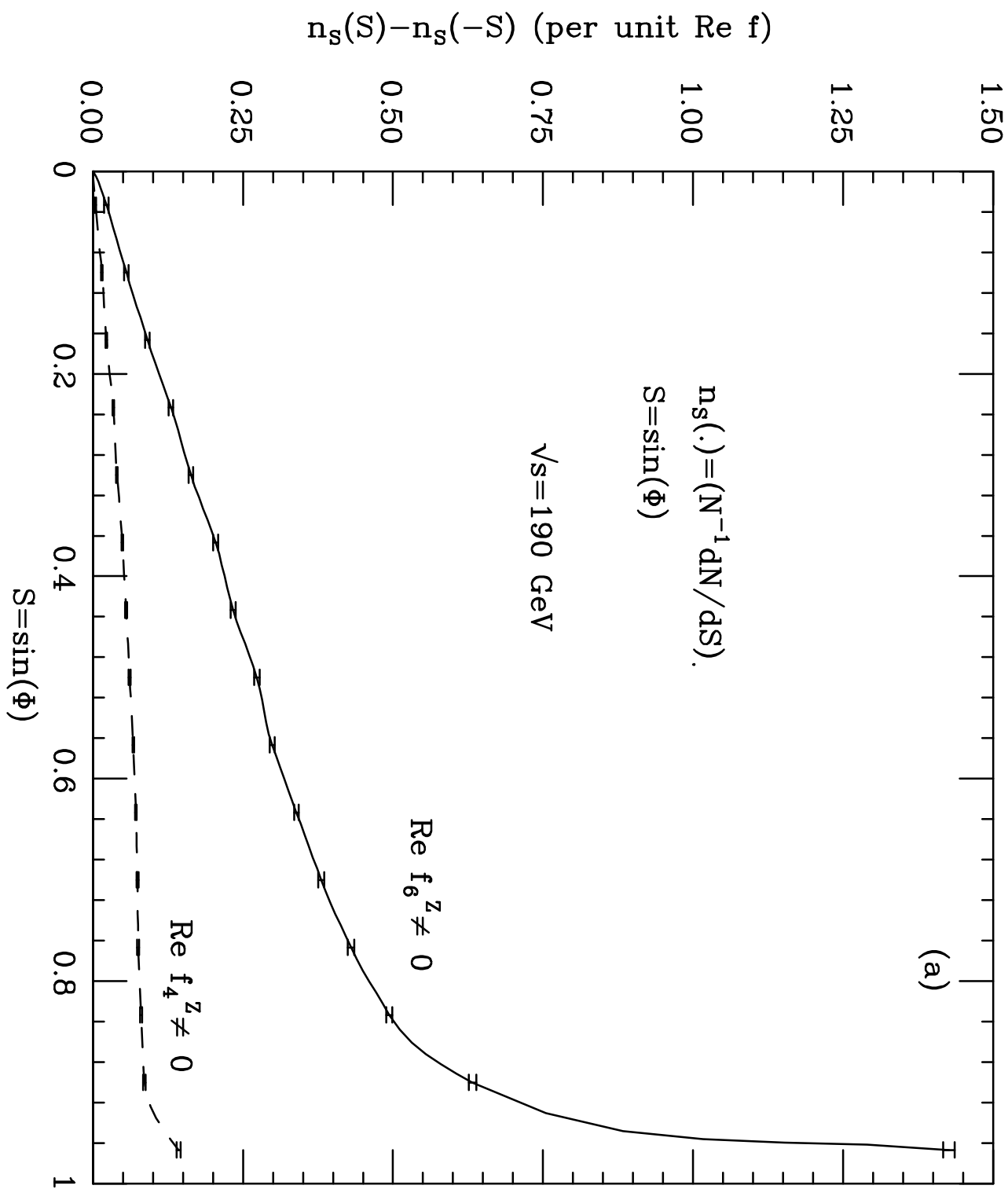
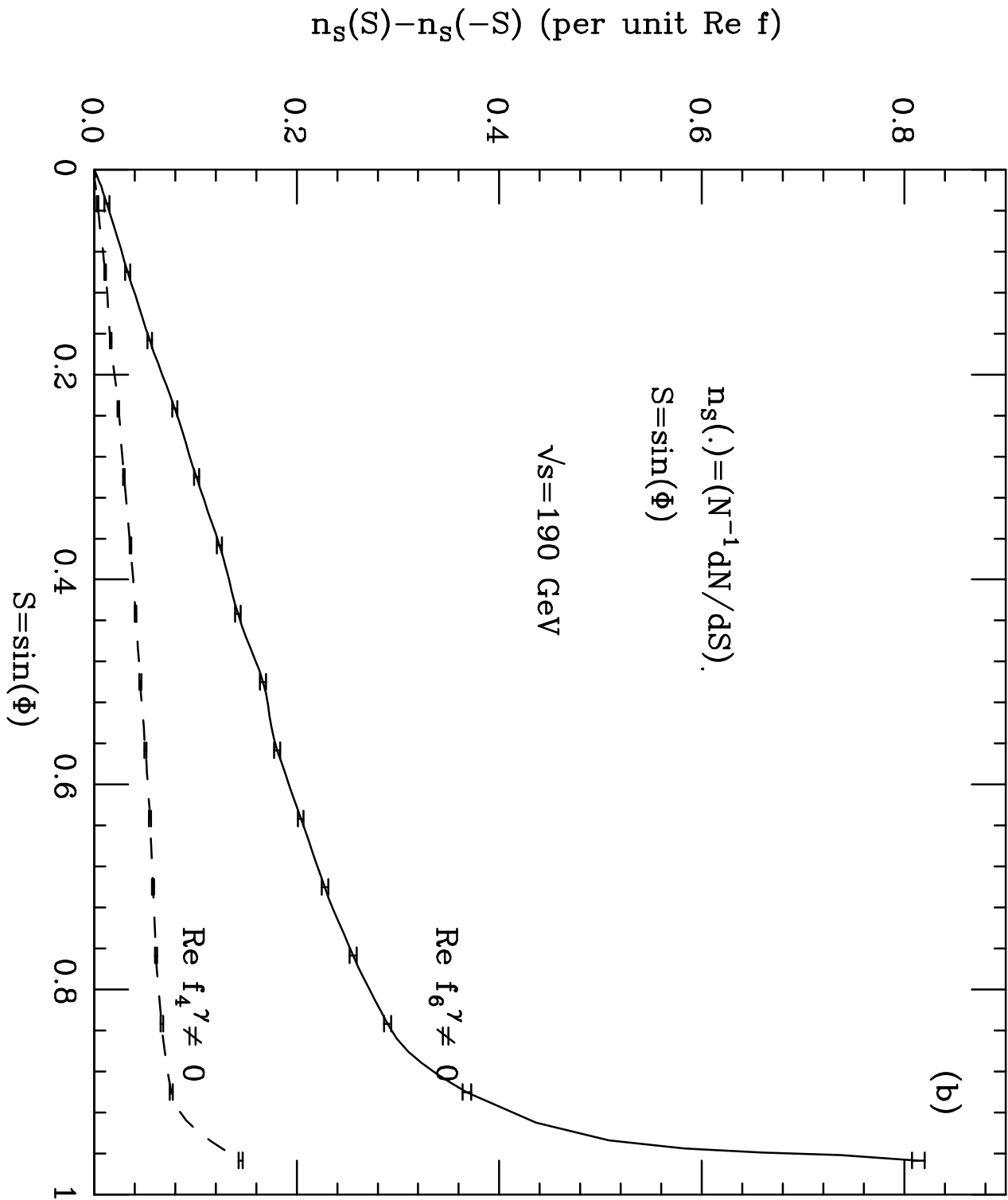
\item[Fig. 9]
(a) Result from Monte--Carlo simulation for the 
distrbution difference $n_S(S)-n_S(-S)$ is 
plotted at $\sqrt{s}=190$ GeV as a function of $S$ per unit of 
$\hbox{Re } f_4^Z$ or $\hbox{Re } f_6^Z$. 
Here $S=\sin\Phi$ and $n_S(S)=N^{-1}dN/dS$.
(b) Same as (a) for the form factors 
$\hbox{Re } f_4^{\gamma}$ and $\hbox{Re } f_6^{\gamma}$. 
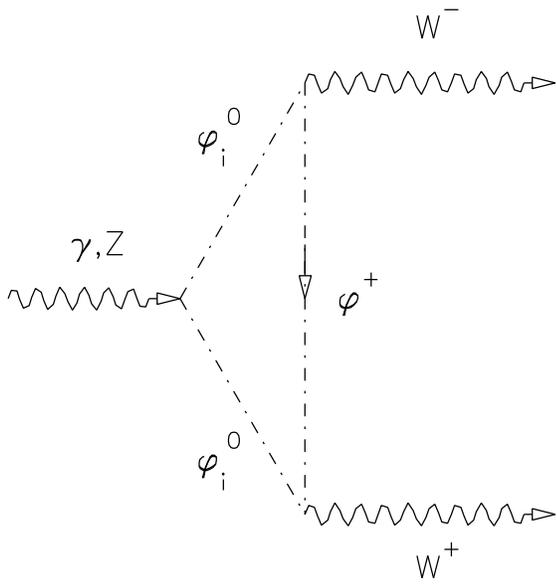
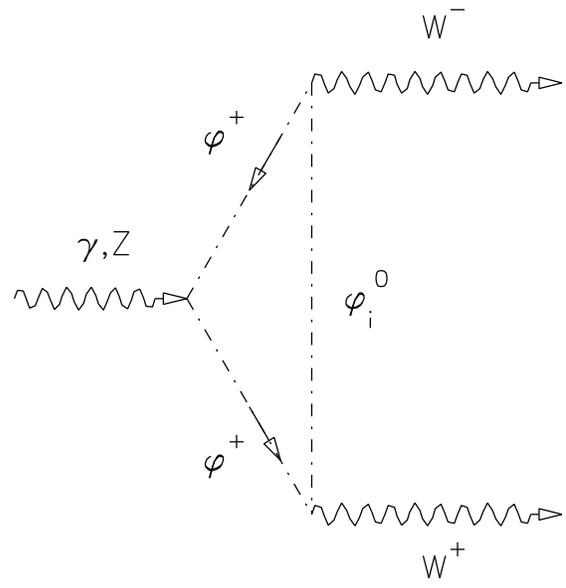
\item[Fig. 10] Diagrams responsible for $f_4^Z$ in the two Higgs doublet model.
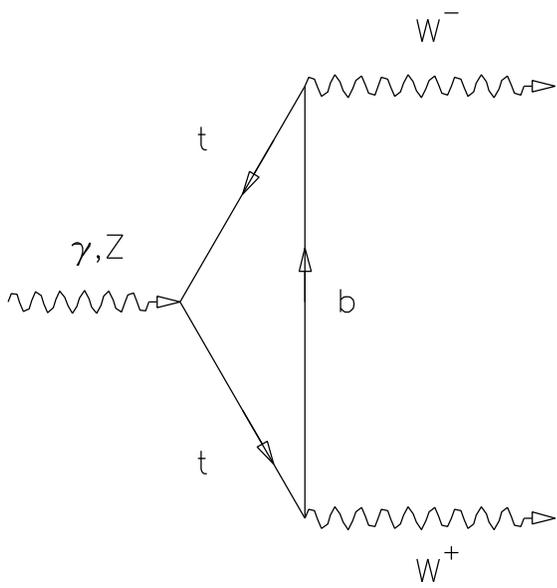
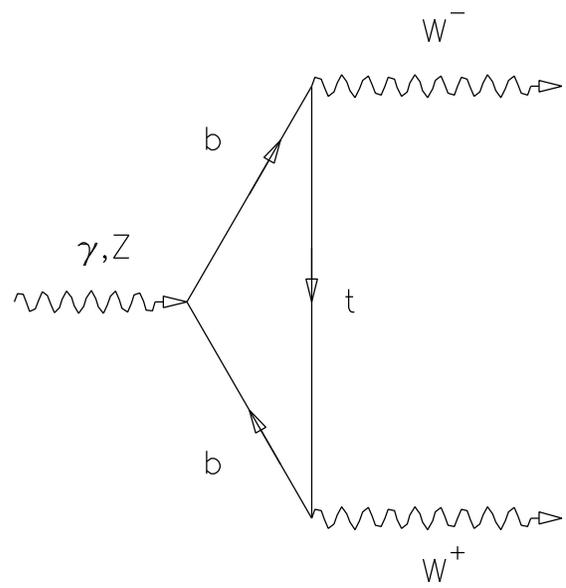
\item[Fig. 11] CP violating Fermion loop diagrams in the Left-Right
symmetric model.
\end{itemize}

\begin{references}
%
\bibitem{cronin} J. H. Christenson, J. W. Cronin, V. L. Fitch, R. Turlay, Phys.
Rev. Lett. {\bf 13} (1964) 138.
%
%
\bibitem{KM}
M. Kobayashi and M. Maskawa, Prog. Theor. Phys. {\bf 49} (1973) 349.
%
\bibitem{hagiwara}
K. Hagiwara, R. D. Peccei, D. Zeppenfeld, and K. Hikasa,
Nucl. Phys. {\bf B282} (1987) 253;
%
\bibitem{ref:Gaemers}
K.J.F. Gaemers and G.J. Gounaris, Z. Phys. {\bf C1}, (1979) 259.
%
\bibitem{gounaris}
G. Gounaris, D. Schildknecht, and F. M. Renard, Phys. Lett. {\bf B263} (1991)
291.
%
\bibitem{othereewwcp}
M. B. Gavela, F. Iddir, A. Le Yaouanc,
L. Oliver, O. Pene and J. C. Raynal,
Phys. Rev. {\bf D39}, 1870 (1989);
A.~Bilal, E. Mass\'o, and A. De R\'ujula, 
Nucl. Phys. {\bf 355}, 549 (1991).
%
\bibitem{r:cphiggs}
D. Chang and W.--Y. Keung, Phys. Lett. {\bf B305}, 261 (1993).
%
\bibitem{r:hcpodd}
D. Chang, I. Phillips, and W.--Y. Keung, 
Phys. Rev. D{\bf 48}, 3225 (1993), or
CERN preprint
CERN--TH.6814/93 (1993); or hep-ph/9303226.
%
\bibitem{r:eett}
D. Chang, I. Phillips,  and W.--Y. Keung, 
Nucl. Phys. {\bf B408}, 286 (1993); Erratum, ibid., B429 (1994) 255;
or  hep-ph/9301259.
%
\bibitem{r:othereett}
J. Korner, J. P. Ma, R. Munch, O. Nachtmann and R. Schopf,
Zeit. Phys. {\bf C51}, 447 (1991);
W. Bernreuther and O. Nachtmann, Phys. Rev. Lett. {\bf 63}, 2787
(1989);
W. Bernreuther, O. Nachtmann, P. Overman and T. Schroder,
Heidelberg Preprint, HD-THEP-92-14~(1992);
W. Bernreuther, J.P. Ma, and T. Schr\"oder,
Heidelberg preprint, HD--THEP--92--30~(1992);
B. Grzadkowski, CERN preprint CERN--TH.6806/93 (1993).
%
\bibitem{r:CPjet}
J.F. Donoghue and G. Valencia, Phys. Rev. Lett. {\bf 58},
451 (1987);
J.F. Donoghue, B. Holstein and G. Valencia, Phys. Lett. {\bf B178}, 319 
(1986);
G. Valencia and A. Soni, Phys. Lett. {\bf B263}, 517 (1991);
J. Korner, J. P. Ma, R. Munch, O. Nachtmann and R. Schopf, 
Zeit. fur Phys. {\bf C50}, 447 (1991);
A. Brandenburg, J. P. Ma, R. Munch and O. Nachtmann,
Zeit. fur Phys. {\bf C51}, 225 (1991);
A. Brandenburg, J. P. Ma and O. Nachtmann,
Zeit. fur Phys. {\bf C55}, 115 (1992).
%
\bibitem{r:ckl}
D. Chang, W.--Y. Keung and J. Liu, 
Nucl. Phys. {\bf 355}, 295 (1991).
%
\bibitem{r:modification}
Note that our transformation is slightly different from that of 
Ref.\cite{hagiwara} because of the change in convention mentioned earlier.
The present form of transformation results in simpler transformation for the 
density matrix also.
%
\bibitem{peskin}
C. R. Schmidt and A. E. Peskin, Phys. Rev. Lett. {\bf 69}, (1992) 410.
%
\bibitem{r:soni}
A. Soni and R. M. Xu, Brookehaven preprint BNL-48160 ITP-SB-92-54.
%
\bibitem{gounaris2}
G. Gounaris, J. Layssac, G. Moultaka, and F. M. Renard, 
Montpellier preprint PM/92-37 THES-TP 92/15(1992).
%
\bibitem{r:He}
X. G. He, J. P. Ma, and B. H. McKellar, University of Melbourne preprint,
UM-P-92/75.
%
\bibitem{pdb}
{\em Review of Particle Properties}, Phys. Rev. D {\bf 45} (1992) page V.13
%
%
\bibitem{r:Fab} E. Christova and M. Fabbrichesi, CERN preprint, 
CERN.6751/92. 
%
\end{references}
\end{document}